\documentclass{emulateapj}
\slugcomment{{\sc Accepted to ApJ:} May 9, 2010} 

\def\msun{{\rm ~M}_{\odot}}

\def\mdot{\dot M}
\def\mpy{{\rm ~M}_{\odot} {\rm ~yr}^{-1}}

\begin{document}

\title{The {\em LISA} Gravitational Wave Foreground: A Study of Double White Dwarfs}

 \author{Ashley J. Ruiter,\altaffilmark{1,2,3} Krzysztof Belczynski,\altaffilmark{4,5,6} 
Matthew Benacquista,\altaffilmark{7} Shane L. Larson\altaffilmark{8} Gabriel Williams\altaffilmark{7}}

\email{ajr@mpa-garching.mpg.de, kbelczyn@nmsu.edu, benacquista@phys.utb.edu, s.larson@usu.edu, gabriel.j.williams@gmail.com}

\altaffiltext{1}{Max-Planck-Institut f{\"u}r Astrophysik,
  Karl-Schwarzschild-Str. 1, 85741 Garching, Germany}
\altaffiltext{2}{Harvard-Smithsonian Center for Astrophysics, 60 Garden St.,
  Cambridge, MA 02138 (Predoctoral Fellow)}
\altaffiltext{3}{New Mexico State University, Dept. of Astronomy,
  1320 Frenger Mall, Las Cruces, NM 88003}
\altaffiltext{4}{Los Alamos National Laboratory, CCS-2/ISR-1 Group}
\altaffiltext{5}{Oppenheimer Fellow}
\altaffiltext{6}{Astronomical Observatory, University of Warsaw, Al.
  Ujazdowskie 4, 00-478 Warsaw, Poland}
\altaffiltext{7}{Center for Gravitational Wave Astronomy, The University of Texas at 
  Brownsville, 80 Fort Brown, Brownsville, TX 78520}
\altaffiltext{8}{Department of Physics, Utah State University, 4415 Old Main Hill, Logan, UT 84322}

\begin{abstract} 

Double white dwarfs are expected to be a source of confusion-limited noise for the future
gravitational wave observatory {\em LISA}. In a specific frequency range, this `foreground
noise' is predicted to rise above the instrumental noise and hinder the detection of
other types of signals, e.g., gravitational waves arising from stellar mass objects
inspiraling into massive black holes.  In many previous studies only
detached populations of compact object binaries have been considered
in estimating the {\em LISA} gravitational wave foreground signal.
Here, we investigate the influence of compact object detached and Roche-Lobe
Overflow Galactic binaries on the shape and
strength of the {\em LISA} signal. 
Since $ > 99$\% of remnant binaries which have orbital periods within the
{\em LISA} sensitivity range are white dwarf binaries,  we consider only
these binaries when calculating the {\em LISA} signal.  We find that
the contribution of RLOF binaries to the foreground noise is negligible at
low frequencies, but becomes significant at higher frequencies, 
pushing the frequency at which the foreground noise drops
below the instrumental noise to $> 6$ mHz.  
We find that it is important to consider the population of mass 
transferring binaries in order to obtain an accurate assessment of 
the foreground noise on the {\em LISA} data stream.  
However, we estimate that there still exists a sizeable number ($\sim 11300$) 
of Galactic double white dwarf binaries which will have a signal-to-noise ratio $> 5$, 
and thus will be potentially resolvable with {\em LISA}. 
We present the {\em LISA} gravitational wave signal from the Galactic
population of white dwarf binaries, show the most important
formation channels contributing to the {\em LISA} disc and bulge populations
and discuss the implications of these new findings.

\end{abstract}


\section{Introduction}
To obtain important information about a variety of astrophysical 
sources, Astrophysics must move toward observations beyond the electromagnetic spectrum.  
Several ground based gravitational radiation (GR)
observatories are already working at full efficiency and collecting data (e.g., LIGO,
\citet{Abb07}).  
These detectors are sensitive in a high GR frequency regime
(above a few tens of Hz), where signals are expected to come from
mergers of compact stellar mass objects like neutron stars (NSs) and black
holes (BHs).  However, these types of events are infrequent, their rate
estimates are burdened with heavy uncertainties \citep[e.g.,][]{Bel02,Kal04,Osh08} 
and it is not clear whether any merger detections are likely given the 
current level of instrument sensitivity \citep{LSC09}.  The future generation of GR
detectors includes {\em LISA}, the Laser Interferometer Space Antenna 
\citep[][and references therein]{Dan96,Hug06}, a joint ESA and NASA mission.  It is a
space-based all-sky interferometer which will be sensitive to lower
frequencies\footnote{For a circular binary, $f_{\rm gr}$ [Hz] $=2/$(orbital period) [s].} 
($\sim 10^{-4} - 0.1$ Hz) than are ground-based GR detectors.  
The different frequency regime will allow for the observations of sources 
of a different nature.  The most promising sources for {\em LISA} are mergers of
supermassive and/or intermediate-mass black holes \citep[i.e.,][]{Ses05}, extreme 
mass ratio inspirals \citep[EMRIs; e.g.,][]{Gai07} of stellar-mass objects into supermassive 
black holes, and stellar mass compact remnant binaries \citep[see e.g.,][]{HBW90,BLT07}.

At first, it was believed that contact W UMa systems \citep{Mir65}
would dominate the GR signal in the low frequency range, but later work 
\citep[e.g.,][]{EIS87,LPP87,HBW90} has shown that close double white
dwarf (WD) binaries are more important at low frequencies, and are a guaranteed 
source of gravitational waves for {\em LISA}.    
The gravitational wave signal arising from double WDs in the Galaxy will put 
constraints on the star formation history of the Milky Way (MW), as well as the scale height 
and shape of the thick disc and Galactic bulge \citep{BH06}.  

The Galactic double WDs will contribute to the {\em LISA} signal as both unresolved and 
resolved sources.  At low frequencies ($\lesssim 3$ mHz; pertaining to binaries with
orbital periods larger than $11$ minutes), predicted binary
numbers are so high that most of the GR signals from individual systems will 
be unresolved, and they will form a confusion-limited `foreground' noise.  Within that
frequency range, only GR sources that are relatively close to the detector
or are strong GR emitters will stand out above the foreground noise.  At higher
frequencies ($\gtrsim 3$ mHz; orbital periods $< 11$ minutes), the numbers 
of double white dwarfs are relatively small and thus they are expected to be resolved, 
offering an opportunity to study uncertain parameters associated with binary evolution 
and WD physics (e.g., possible progenitors of SNe Ia and R CrB stars).        
{\em LISA} will be mostly sensitive to Galactic stellar-mass binaries, however a small 
contribution due to extragalactic systems was also predicted \citep{HBW90,KP98,FP03}.  
The gravitational wave radiation 
from coalescing double WDs was investigated using smoothed particle hydrodynamics 
by \citet{Lor05}.  It is expected that the signal arising 
from merging double WDs would be detectable prior to the
coalescence, though would not contribute significantly to the {\em LISA} 
gravitational wave signal overall, as the Galactic double WD merger event rate is predicted 
to be as low as $\sim 1$ per century \citep{NYP01}.   

Among the objects which are expected to be important verification sources
for {\em LISA} are the compact AM Canum Venaticorum (AM CVn) binaries
(see \citet{Roe07} for gravitational wave strain amplitude
estimates of five known AM CVn binaries).  AM CVn systems
are a sub-class of cataclysmic variables with $\lesssim
1$ hour orbits in which a WD accretes matter via Roche-Lobe overflow
(RLOF) from what is believed to be a helium-rich white dwarf or a (semi-degenerate) helium
star \citep{Sma67,War95,Nel01}.  These close binaries are faint
and thus not readily detected electromagnetically, but are expected to
be observable in gravitational waves (25 AM CVn systems 
have been confirmed to
date)\footnote{http://www.astro.ru.nl/\~{}nelemans/dokuwiki/doku.php?id\\
=verification\_{}binaries:am\_{}cvn\_{}stars}.

Besides double WD binaries, other close compact binaries including NSs 
and BHs have been studied by other groups \citep[e.g.,][]{NYP01,Coo04}.  Some few tens of these systems 
with orbital periods within the {\em LISA} sensitivity range are known in the Galaxy.
There are also $\sim 8$ ultracompact X-ray
binaries, which consist of a WD transferring matter through RLOF to a
NS on $\sim 40$ minute orbits \citep{WC04}.  Double NSs are
also known, however only one such system is observed with a period (of
2.4 hrs) within the {\em LISA} sensitivity range \citep{Bur03}.  Close
binaries with BHs (like BH-BH, BH-NS, BH-WD) have yet to be
observed, but they are predicted to populate the Galaxy in smaller
numbers than systems with NSs and/or WDs.  

Previous work on Galactic compact binary populations in context of
{\em LISA} has been done with the exclusion of RLOF systems \citep{HBW90,PP98,BDL04,TRC06}
or with the entire RLOF $+$ detached populations combined \citep{NYP01,Edl05}.  
In addition, \citet{HB00} studied AM CVn systems separately using 
an analytical approach, finding no significant 
increase in the confusion-limited noise for space-based GR detectors with the addition of RLOF 
binaries, while \citet{NYP04} calculated the GR amplitude of 
resolved WD binaries using population synthesis methods (both detached
and AM CVn systems), finding that {\em LISA} would resolve 
a total of $22000$ Galactic double WDs ($\sim 11000$ detached, $11000$
AM CVn).  However, a more recent study which takes into account the local space
density of AM CVn stars predicts that only $\sim 10^{3}$ AM CVn
binaries will be resolved with {\em LISA} \citep{RNG07}.  This
estimate is still $\sim 2$ orders of magnitude above the current 
number of AM CVn systems which have so far been detected.  

In modeling our stellar population, we use an updated population synthesis method 
with recent results on mass transfer/accretion in compact object binaries, that is significantly 
different and more complex than the methods used in previous calculations.  
In this study, we assess the importance of detached and RLOF compact
binary populations in the Galactic disc and bulge on the 
overall {\em LISA} sensitivity.  In \S\, 2 we describe our binary modeling 
techniques and signal calculations, in \S\, 3 we show the physical properties (masses, orbital 
periods) of our double WD population, and show the most important evolutionary 
channels which contribute to the overall {\em LISA} signal arising from
the four Galactic sub-populations (detached disc, RLOF disc, detached
bulge, RLOF bulge).  In \S\, 4 we show the {\em LISA} GR signals, estimate the
likelihood of source resolvability, and comment on the detectability 
and selection effects of the potentially resolvable systems within 
our Galaxy.  In \S\, 5 we compare our results  
with those of previous studies, and in \S\, 6 we give a summary of our
findings. 

\section{Model Description}

\subsection{Population synthesis model}

In this study, the combined Galactic population of RLOF and detached 
compact remnant binaries with orbital periods spanning the {\em LISA} band is considered. 
We use the {\tt StarTrack} population synthesis
code described in detail in \citet{Bel08} to evolve 
primordial binaries within the Galaxy from the Zero-Age Main Sequence
(ZAMS) and calculate their properties.  The code has undergone
a number of important updates since \citet{Bel02}, including detailed 
mass transfer calculations.  Recent
results on mass accumulation in context of ultracompact X-ray binary
formation are given in \citet{BT04}, while for double
WD binaries in \citet{BBR05}.  
 
We evolve the field population of single and binary stars ($50$ \%
binarity) with solar-like metallicity ($Z=0.02$).
Single star evolution is followed from the ZAMS employing 
modified analytic formulae and evolutionary tracks from \citet{HPT00}.  
For the evolution of binary stars, the
calculations also include full orbital evolution with tidal
interactions, magnetic braking, supernova natal kicks and 
gravitational radiation emission \citep[see][for formulae]{Bel08}.  
ZAMS masses ($M_{\rm ZAMS}$) span the mass range $0.08-100
\msun$.
Single stars and binary primaries ($M_{\rm a}$) are drawn from a 3-component broken power
law initial mass function \citep{KTG93}, and secondary masses ($M_{\rm b}$) are obtained
from a flat mass ratio distribution $q=$secondary/primary
\citep[e.g.,][]{Maz92}. 
We note as well that binary interactions leading
to mass transfer events may alter 
the course of evolution for stars of a given initial mass.  
Initial orbital
separations span a wide range up to $10^{5}$ R$_{\odot}$ and are
drawn from a distribution which is flat in the logarithm, while
initial eccentricities are taken from a 
thermal-equilibrium eccentricity distribution \citep[see][]{Abt83,DM91}.
Progenitors of all binaries are 
initially formed on eccentric orbits.  Tidal interactions between binary 
components circularize orbits before the first RLOF in a given progenitor system occurs.  
However, for a small fraction (10\%) of double WD progenitors that are
found initially on very eccentric 
orbits, the first RLOF is encountered when the orbit is still 
eccentric.  In such a case we assume an instant circularization at periastron 
($a_{\rm new}=a_{\rm old}\,(1-e_{\rm old})$ and $e_{\rm new}=0$), and follow with the RLOF 
calculation.  At the time of double WD formation all of the orbits are
circular.  Eccentric WD binaries are expected to arise from dynamical interactions
in globular clusters \citep{Ben01}, and though we do not consider them
here could provide a unique opportunity for learning about WD
structure with {\em LISA} \citep{WVK08}.

For the population of Galactic disk binaries, we use a continuous star
formation rate for 10 Gyr, while for the bulge population we use a
constant star formation rate for the first Gyr, and no star formation
thereafter (entire MW age is 10 Gyr). This translates to a present
Galactic star formation rate of $4$ M$_{\odot}$ yr$^{-1}$ (disc mass $= 4
\times 10^{10}$ M$_{\odot}$ with a constant star formation rate for 10 Gyr
$= 4$ M$_{\odot}$ yr$^{-1}$; see \S\, 2.2), which is within reasonable agreement of the
current Galactic star formation rate estimate of $3.6$ M$_{\odot}$
yr$^{-1}$ \citep{Cox00}. We note however that it has been suggested that
the star formation rate of the MW has been (mostly) decreasing
exponentially with time, only reaching $\sim 3.6 \mpy$ at the current
epoch \citep[][see sect. 2.2]{NYP01,NYP04}; the integrated mass in formed
stars being closer to $8 \times 10^{10}$ M$_{\odot}$.  Since the star
formation history of the MW disc and bulge is not precisely known, we
choose to model the Galaxy with simple star formation histories (1
Gyr-long `continuous burst' (bulge), and constant for 10 Gyr (disc)),
which to first order is a reasonable representation of the global
star formation history of the Galaxy.

The most uncertain phase in close binary evolution affecting the
orbital separation for low- and intermediate- 
mass stars is the common envelope (CE) phase \citep{PY06}.  
Close binaries are expected to go through at least one common envelope event,   
and there are currently a few prescriptions of CE evolution in the literature 
\citep{Web84,NT05,VBP06,Bee07}.  
In our simulations the CE phase is treated using energy balance as
discussed in \citet{Web84},
in which the orbital energy of the binary is diminished at the expense
of the unbinding of the donor envelope.  The post-CE separation  
is governed by the parameters $\lambda$ \citep{deK90}, a function of the donor's
structure, and the highly uncertain $\alpha$, the efficiency with 
which the binary orbital energy is used to unbind the envelope.  
We have chosen to use $\alpha \times \lambda = 1$.  The effects of 
using other CE efficiencies have not been fully explored here, though 
lower CE efficiencies result in closer post-CE binaries, and a higher number 
of stellar mergers.  

At the current age of the Galaxy (10 Gyr) we extract all  
binaries containing two compact remnants: 
WDs, NSs and BHs in the gravitational frequency 
range: $0.0001 - 1$ Hz, which
encompasses the {\em LISA} sensitivity range (orbital periods of 5.6 hr
-- 2s).  Henceforth when we refer to `{\em LISA}
binaries', we are referring to binary systems in our study within this 
GR frequency band.  
As previously stated, we consider only white dwarf binaries 
when calculating the {\em LISA} signal. 
There are 5 types of WDs considered in the evolution: helium (He WD), 
formed by stripping the envelope off a Hertzsprung gap or red giant low-mass star; 
carbon-oxygen (CO WD), formed from progenitors with masses 
$\sim 0.8-6.3 \msun$; oxygen-neon (ONe WD), formed  from progenitors with masses 
$\sim 6.3-8.0 \msun$; hybrid (Hyb WD), having a CO-core and a
He-envelope, formed via the stripping of the envelope from a helium 
burning star, and hydrogen (H WD), formed by stripping the envelope from a 
very low mass ($\lesssim 0.8 M_{\odot}$) main sequence (MS) star.  
In our {\em LISA} GR calculations, we only consider the first four types of WDs, as
hydrogen white dwarfs are more representative of brown
dwarf-like objects and thus we will from now on refer to them as brown
dwarfs.  For the current study, the brown
dwarfs (formed through binary evolution) do not play a large role in contributing to the
{\em LISA} GR signal due to their low mass (see \S\, 5).     
We note however that in older stellar populations, the fraction of 
binaries which contain H WDs becomes more significant \citep[see][for the
calculation of the {\em LISA} signal from MW halo double white dwarfs,
which includes `hydrogen white dwarfs']{Rui09}.  

In compact binaries, gravitational wave emission is a strong source of angular momentum loss.   
The average rate of orbital angular momentum loss due to GR for $e=0$ binaries is 
calculated from \citet{Pet64}
\begin{equation}
\frac{dJ_{\rm gr}}{dt}=\frac{-32}{5}\frac{G^{7/2}\,M_{\rm p}^{2}\,M_{\rm s}^{2}\,\sqrt{M_{\rm p} + M_{\rm s}}}{c^{5}\,a^{7/2}}
\end{equation}
where $G$ is the gravitational constant and $c$ is the speed of light.  
At every time step in our calculations 
the evolving system is checked for RLOF.    
Upon reaching contact (RLOF), we assume that mass-transfer is GR-driven and
the donor ($M_{\rm don}$) mass transfer rate is calculated via: 
\begin{equation}
\mdot_{\rm don} = M_{\rm don} D^{-1} {\,d J_{\rm gr}/\,d t \over J_{\rm orb}}
\end{equation}
where RLOF stability is determined by the parameter $D$ \citep[see i.e.,][]{KK95}
\begin{equation}
D={5 \over 6}+ {1 \over 2} \zeta_{\rm don}-{1-f_{\rm a} \over 3 (1+q)}-
{ (1-f_{\rm a}) (1+q) \beta_{\rm mt}+f_{\rm a} \over q}. 
\end{equation}
$J_{\rm orb}$ is the orbital angular momentum of the binary, $f_{\rm
  a}$ is the fraction of transferred mass 
accreted by the WD of mass $M_{\rm acc}$ ($f_{\rm a}=1$ for stable RLOF), $q \equiv (M_{\rm
  acc}/M_{\rm don})$,  
and $\beta_{\rm mt} =  M_{\rm don}^{2}/(M_{\rm don} + M_{\rm acc})^{2}$. 
The radius mass exponent for the donor $\zeta_{\rm don}$ is 
obtained from stellar models in each time step \citep[see][]{Bel08}
  and is $\sim -0.35$ for WD donors in a phase of stable mass transfer.  
Stable RLOF between two WDs leads to an increase in orbital period where dynamically unstable 
RLOF leads to a merger.  See \citet{Bel08} for a more detailed
description of treatment of mass transfer/accretion phases in {\tt
  StarTrack}.  

\subsection{Calibration and Spatial Distribution of Sources}

In our simulations both detached and RLOF systems have been
distributed about the disc and bulge separately, 
assuming no correlation between position and age of the
system.  For the current study, 
we have neglected the halo population of double WDs, but these systems have been investigated  
in another work \citep{Rui09}. 
The density distribution of the Galactic disc is taken to have the 
following form:
\begin{equation}
   \rho({\bf R,z}) = \frac{N_{d}}{4\pi R_{0}^{2} z_{0}}e^{-R/R_{0}} e^{|-z|/z_{0}}
\label{disc}
\end{equation}
in cylindrical coordinates, and the density distribution of the bulge
has the form:
\begin{equation}
   \rho({\bf r}) = \frac{N_{b}}{4\pi r_{0}^3}e^{{-(r/r_{0})}^2}
\label{bulge}
\end{equation}
in spherical coordinates (bulge scale length $r_0 = \sqrt{x^{2} +
  y^{2} + z^{2}}$).  

$N_{d}$ and $N_{b}$ represent the total numbers of simulated double WDs in the disc and 
bulge, respectively.  
We have calibrated our results using stellar
Galactic disc and bulge masses from \citet{KZS02}, with masses of $4
\times 10^{10}$ M$_{\odot}$ and $2 \times 10^{10}$ M$_{\odot}$, respectively.  
We have chosen $R_0 = 2500~{\rm pc}$ and $z_0 = 200~{\rm pc}$ 
for the disc scale length and scale height while $r_0 = 500~{\rm pc}$
with a radial cut-off of $3500$ pc for the bulge \citep{Nel03}.  
While we note that even a scale height of $500$ pc would be a reasonable 
choice for disc WDs \citep{MS02}, we have chosen to use $200$ pc so
that our results are more readily comparable to those of previous 
studies \citep{NYP01,BH06}.  
   
Our original {\em LISA} double WD binaries which were born directly from {\tt StarTrack} 
($8.4 \times 10^{4}$) were the result of the evolution of $20 \times 10^{6}$ 
ZAMS binaries.  
Since evolving the whole Galactic population of {\em LISA} binaries is time-intensive, an
interpolation  scheme was developed in order to scale our number of {\tt StarTrack} binaries to match
those of the Milky Way bulge and disc by stellar mass.
Probability distribution functions (PDFs) were constructed for
different evolutionary 
channels as a function of the primary and secondary masses and GR frequency.
For RLOF
binaries, it was only necessary to construct PDFs as a function of the
masses since the orbital period, and hence GR frequency, was then
uniquely determined by the masses.  We chose to
construct PDFs (instead of simply scaling the original binaries by a factor of $\gtrsim
300$) so that our systems would span a range of frequencies and masses pertinent to
said formation channel.  The PDFs
were constructed using a kernel density estimator that best replicated the characteristics
of the population~\citep{williams08}. The KDE package for Matlab was used to generate the
specific PDFs. 
This technique was particularly successful for well-populated evolutionary
channels, where we could be sure that the range of masses and frequencies was well
covered.  For some of the least-populated channels, 
the full range of binary masses and frequencies was not covered
sufficiently to yield a smooth extrapolation of the data.  
Thus, apparent clumps of binaries that are comparable in number to our
scaling factors are attributable to an under-sampling of the frequency 
and mass range for these rare evolutionary channels.

\subsection{Lisa Signal Calculations}

The gravitational wave signal from the {\em LISA} data stream will be a time 
delay interferometry (TDI) variable in which the phases of the laser 
signals at each vertex of {\em LISA} are combined with the phases of other 
signals at delayed times in order to reduce the laser phase noise and 
accommodate the varying armlengths of the constellation of the 
spacecraft \citep{RCP04}. At low frequencies, the signal at any 
given vertex can be very well approximated by the Michelson signal:
\begin{equation}
h(t) = \frac{1}{2} h_{ab}\left(\ell_1^{a}\ell_1^{b} - 
\ell_2^{a}\ell_2^{b}\right),
\label{michelson}
\end{equation}
where $h_{ab}$ is the wave metric and $\ell_1$ and $\ell_2$ are the 
unit vectors pointing along the two arms that join at the vertex. 
The transfer 
frequency is the frequency at which point the gravitational radiation 
wavelength becomes comparable to the armlength of {\em LISA}.  For high 
(e.g., above the transfer frequency) GR frequencies, the period of 
the gravitational wave is then less than the time it takes for 
light to propagate between {\em LISA} spacecraft detectors, and the low frequency
approximation breaks down.  The transfer frequency is given by 
$f_{*} = c/(2 \pi L) \approx 0.01$ Hz, where $L$ is the {\em LISA} 
arm-length taken to be $L = 5\times 10^9~{\rm m}$.  
At frequencies above or near $f_{*}$ a more accurate representation of the 
{\em LISA} signal is given by the \emph{rigid adiabatic approximation} 
\citep{RCP04}, but detailed analysis 
\citep{VW04} have shown the low-frequency approximation 
should be adequate for most tasks at frequencies below 
$\sim 30~{\rm mHz}$.  In this work, we have used the rigid adiabatic
approximation to calculate the {\em LISA} signal for sources with
frequencies above 0.003 Hz.  

Space-borne gravitational wave observatories like {\em LISA} are constantly
in motion, changing their relative speed and aspect with respect to
astrophysical sources on the sky.  The sensitivity of the observatory
to different gravitational wave polarizations as a function 
of sky position is reflected
in the antenna beam patterns, which have the highest gain in the
direction perpendicular to the plane of the interferometer.  As the
observatory moves in its orbit, the gravitational wave signal is
modulated in amplitude, frequency, and phase.  Frequency (Doppler)
modulation arises from the relative motion of the observatory and the
source, amplitude modulation arises from the sweep of the anisotropic
antenna pattern on the sky, and phase modulation results from the
detector's changing response to the gravitational wave polarization
state. The Michelson signal can be calculated using Eq.\ (\ref{michelson})
by either holding the source fixed and putting all the motion of the 
detector into $\ell_1$ and $\ell_2$ \citep{RCP04}, 
or by holding the detector fixed and putting the motion into the 
source \citep{Cut98}. Both approaches yield the same Michelson signal 
in the low-frequency limit:
\begin{equation}
h(t) = \frac{\sqrt{3}}{2} A(t) \cos{\left[\int^t2\pi f(t^{\prime}) 
dt^{\prime} + \varphi_{p}(t) + \varphi_D(t) + \varphi_0 \right]}
\end{equation}
where $A(t)$ is the amplitude modulation, $f = 2/P_{\rm orb}$ (where $P_{\rm orb}$ 
is the binary orbital period) is the gravitational wave frequency 
for systems with zero eccentricity, and $\varphi_{0}$ is the initial phase of 
the wave at $t = 0$.  The polarization phase, $\varphi_{p}(t)$, represents the phase 
modulation. The Doppler phase, $\varphi_D(t)$, gives the frequency 
modulation and the amplitude modulation is described by: $A(t) = \sqrt{(A_+F^+)^2+(A_{\times}F^{\times})^2}$. 
The sensitivity factors, $F^{+}$ and $F^{\times}$ are complicated 
functions of the position angles and orientation of the binary, as 
well as the position of {\em LISA} in its orbit 
\citep[see][for specific forms of these functions]{Cut98,RCP04}. 
The plus and cross polarization amplitudes are given by: 
\begin{eqnarray}
A_+ & = & 2 \frac{G^{5/3}}{c^4d}{\cal M}^{5/3} (\pi \,f)^{2/3}(1+\cos^2{i}) \\
A_{\times} & = & -4 \frac{G^{5/3}}{c^4d}{\cal M}^{5/3} (\pi \,f)^{2/3} 
\cos{i},
\end{eqnarray}
where $d$ is the distance to the binary, $i$ is the angle of inclination, 
and the chirp mass is ${\cal M} = \left(M_{\rm p}M_{\rm s}\right)^{3/5}/\left(M_{\rm p}+M_{\rm s}\right)^{1/5}$. 
We have used the approach of \citet{RCP04} to 
calculate the timestreams, which are then added to 
produce the total observatory data stream, which is then Fourier 
transformed to produce the illustrated frequency domain 
representation of the double WD Galactic foreground.

The {\em LISA} instrument noise is simulated by assuming the power spectral density of the noise is
made up of position (or shot) noise given by \citet{Cor02}:
\begin{equation}
S_{\rm np} = 4.8 \times 10^{-42}~{\rm Hz}^{-1},
\end{equation}
and an acceleration noise (converted to strain) given by:
\begin{equation}
S_{\rm na} = 2.3\times 10^{-40}\left(\frac{10^{-3}~{\rm Hz}}{f}\right)^4~{\rm
Hz}^{-1}.
\end{equation}
These separate components are combined according to:
\begin{equation}
S_{\rm n} = 4 S_{\rm np} + 8S_{\rm na}\left(1+\cos^2{\left(f/f_{*}\right)}\right),
\end{equation}
where $f_{*}$ is the transfer frequency.  We roll off the acceleration below $f_{\rm min}=10^{-5}~{\rm Hz}$,
so that $S_{\rm na}(f \le f_{\rm min}) = S_{\rm na}(f_{\rm min})$. In reality, the
{\em LISA} noise will probably not follow this simple power law all the way down to our
choice of $f_{\rm min}$, but will begin to rise at a higher frequency below 0.1
mHz.

\section{Results}

Out of our total population of compact remnant binaries 
with gravitational wave frequencies within the {\em LISA} sensitivity range, 
we note that $\sim 76$\% are double WDs, $\sim 24$\% are binaries involving a
white dwarf and a brown dwarf (WD-BD),
while other types of binaries make up
the rest (i.e., $<0.5$ \% of binaries are NS-WD binaries; less than $0.1$ \%
are NS-NS).  Specific predictions for double compact objects with NSs and
BHs are discussed in \citet{BBB08}.  
  
Relevant for the current study of double white dwarfs, 
we find $N= 34.2 \times 10^{6}$ {\em LISA} binaries in the
disc$+$bulge population (He, CO, ONe and hybrid WDs)\footnote{When
  binaries involving brown dwarfs are considered, $N= 45 \times 10^{6}$.}  .  
Out of this population $N_{d} = 24.8 \times 10^{6}$ (73 \%) 
and $N_{b} = 9.4 \times 10^{6}$ (27 \%).  We find $8.6 \times 10^{6}$ (25
\% of total) of disc binaries are detached and $16.2 \times 10^{6}$
(47 \% of total) of disc binaries are RLOF, while for
the bulge $3.9 \times 10^{6}$ (12 \% of total) are detached and $5.4 \times 10^{6}$
(16 \% of total) are RLOF.  
The total number of detached binaries among the total disc$+$bulge
{\em LISA} population is $12.5 \times 10^{6}$ (37 \%; $8.6 \times
10^{6}$ and $ 3.9 \times 10^{6}$
for the disc and bulge respectively) while for RLOF, often neglected in
previous studies, the number is $21.6 \times 10^{6} $ (63 \%; 
$16.2 \times 10^{6}$ and $ 5.4 \times 10^{6}$ in the disc and bulge, respectively).
Within our RLOF population, $> 99$ \% ($21.4 \times 10^{6}$) are AM CVn-like systems (a WD
accreting from a helium or hybrid WD), the majority of which are 
COWD-HeWD binaries.  

The number of AM CVn systems which we find to currently exist in the
MW can be compared to the estimate of \citet{Nel01}.  However, We 
remind the reader that in our population synthesis, although we do form AM CVn
stars through the helium star channel, we only include
{\em binaries consisting of two degenerate objects} in the final results \citep[see][for a
description of AM CVn formation channels]{Nel01}.  Thus, we cannot
perform a direct comparison with the AM CVn population of \citet{Nel01}.  \citet{Nel01} find
the current number of Galactic AM CVn (double WD channel only) to range
between ($0.2 - 49$) $\times 10^6$, depending on the assumed
effective tidal coupling (see their \S\, 3.4), 
and they find corresponding space
densities of (0.4 - 1.7) $\times 10^{-4}$ pc$^{-3}$, respectively.  
Thus, we find that our
AM CVn number estimate ($21.4 \times 10^{6}$) is in reasonable agreement with that of
\citet{Nel01} if in fact the double WD evolutionary pathway 
is an efficient channel for creating these systems.  

In the study of \citet{RNG07}, it was determined that 
population synthesis studies were overestimating the space
densities of AM CVn binaries, and that the true local space density 
is $\rho_{0} = 1-3 \times 10^{-6}$ pc$^{-3}$, based on
spectroscopic observations of 6 SDSS-I AM CVn stars.  Adopting
the disc-like density distribution as discussed in \S\, 2, we find
that the local space density of AM CVn binaries from our population
synthesis model is $ 2.3 \times 10^{-5}$ pc$^{-3}$,\footnote{For a
  spherical volume with a radius of 200 pc.} which is still an order
of magnitude above the most up-to-date observational estimate.  It is
becoming more clear that mass transfer between two WDs may not be as
efficiently sustained as has been assumed in most population synthesis
studies, and that a large fraction of these systems should merge upon 
reaching contact rather than enter the AM CVn phase \citep{MNS04}.  

\subsection{Formation Channels}

We note that in our calculations, the total population of WD binaries 
with GR frequencies between $1 \times 10^{-4} - 1$ Hz comprises roughly 6\% 
of the \textit{total} {\tt StarTrack} Galactic population of double WDs.  While 
we include all RLOF double WD systems that currently exist in the MW 
in our study, only 2.3\% of Galactic detached double WD systems are accounted 
for, given the orbital period cut-off of $\sim$ 5.6 hrs.  
In Tables 1, 2, 3 and 4, we show the evolutionary history (formation
channels) of the most common {\em LISA} 
double WDs.  We indicate the contribution (percentages) of said channel to
the population of {\em LISA} WD binaries from that particular Galaxy
component (disc, Tables 1 \& 2, or bulge, Tables 3 \& 4).  The
characters and numbers in parentheses in the central column of the tables represent 
the formation channel histories, and recount various stages of stellar evolution
of the primary and secondary star progenitors 
(see Table 1 caption for description). 

In our formation channel notation, for a `CO-He' binary 
the CO white dwarf is the first formed WD, which is not necessarily
the more massive WD (thus is not necessarily the primary star).   
In general, the detached systems have evolved from progenitors in which the stars 
initially have comparable masses,\footnote{Many RLOF systems originate
  from systems in which the initial primary star was much more massive
  than the initial secondary, leading to a CE.} 
thus first RLOF mass transfer is more often dynamically stable and the binary does not undergo 
a CE at this stage (such as He-He detached double WDs).  Typically the detached
double WD progenitors go through only one CE event, 
where as in many cases the RLOF binaries undergo two CE events.

We wish to point out that in
the case of several RLOF binaries (e.g., those involving hybrid donors), there
is a short-lived detached phase preceding the long-lived RLOF phase.
For example, in the evolutionary history of a COHyb-R1 binary, the
progenitor system only spends $\sim 10^{2}$ Myr as a detached CO-Hyb
WD after the second CE phase before driven to contact via GR.  Once
contact is reached however, the binary will
spend a long time (Gyr) as a stable RLOF system.  
This explains why the COHyb-R1 channel shows up in Table 2, but its
predecessor COHyb detached channel is absent in Table 1
(COHyb-D1 binaries only account for $\sim 1 \%$ of the disc
formation channels for {\em LISA} double WDs).  

\begin{figure*}
\includegraphics[width=.9\textwidth]{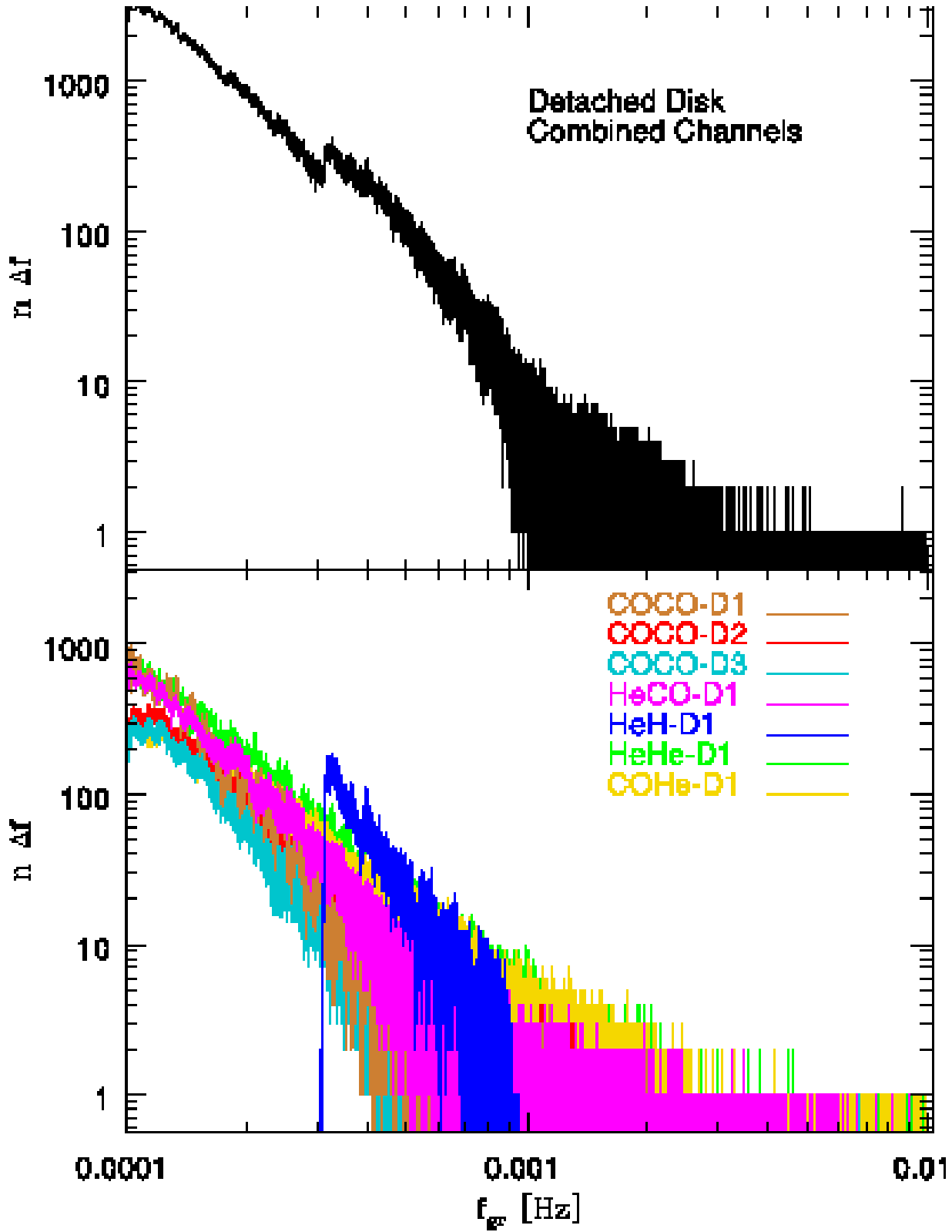}
\caption{Number density ($n = (dN/df)$; $\Delta\,f$ is the size of a resolvable 
bin for a 1 year observation time, 1/$T_{obs} = 30$ nHz) of the
most prominent detached {\em LISA} double WD evolutionary channels for the
Galactic disc.  Some channels from Table
1 have been left out for clarity, and we additionally show some of the
more prominent {\em LISA} binary channels involving brown dwarfs
(neglected in the signal calculations).  Note that as the frequency
increases, so does the concentration of resolvable bins.}
\label{dndfdisdet}
\end{figure*}

\begin{figure*}
\includegraphics[width=.9\textwidth]{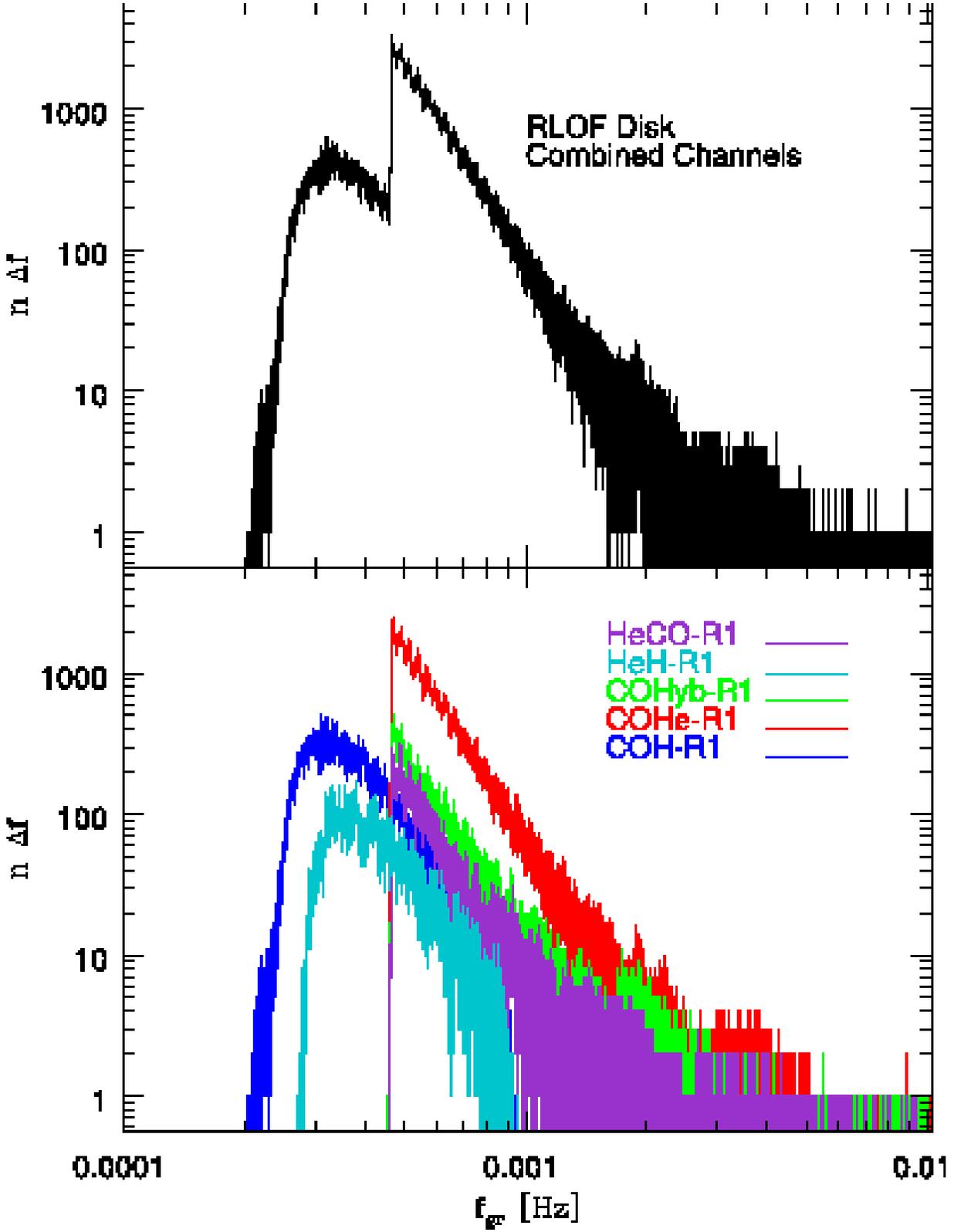}
\caption{Number density of the most prominent RLOF {\em LISA} double WD 
evolutionary channels for the Galactic disc.  All binaries with GR 
frequencies $\lesssim 0.45$ mHz involve binaries with brown dwarf   
donors.}
\label{dndfdisrlo}
\end{figure*}

\begin{figure*}
\includegraphics[width=.9\textwidth]{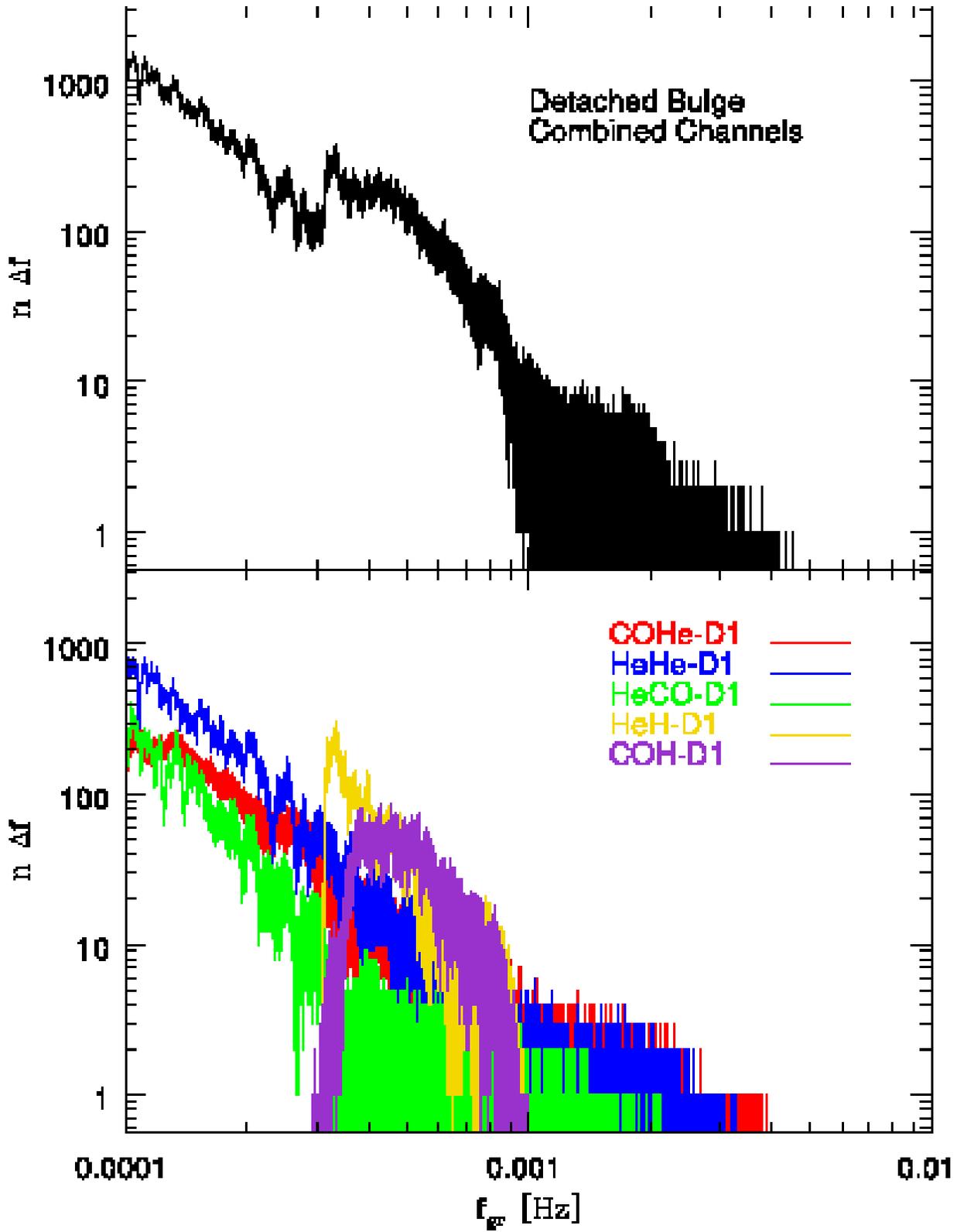}
\caption{Number density of the most prominent detached {\em LISA} double WD 
evolutionary channels for the Galactic bulge.}
\label{dndfbuldet}
\end{figure*}

\begin{figure*}
\includegraphics[width=\textwidth]{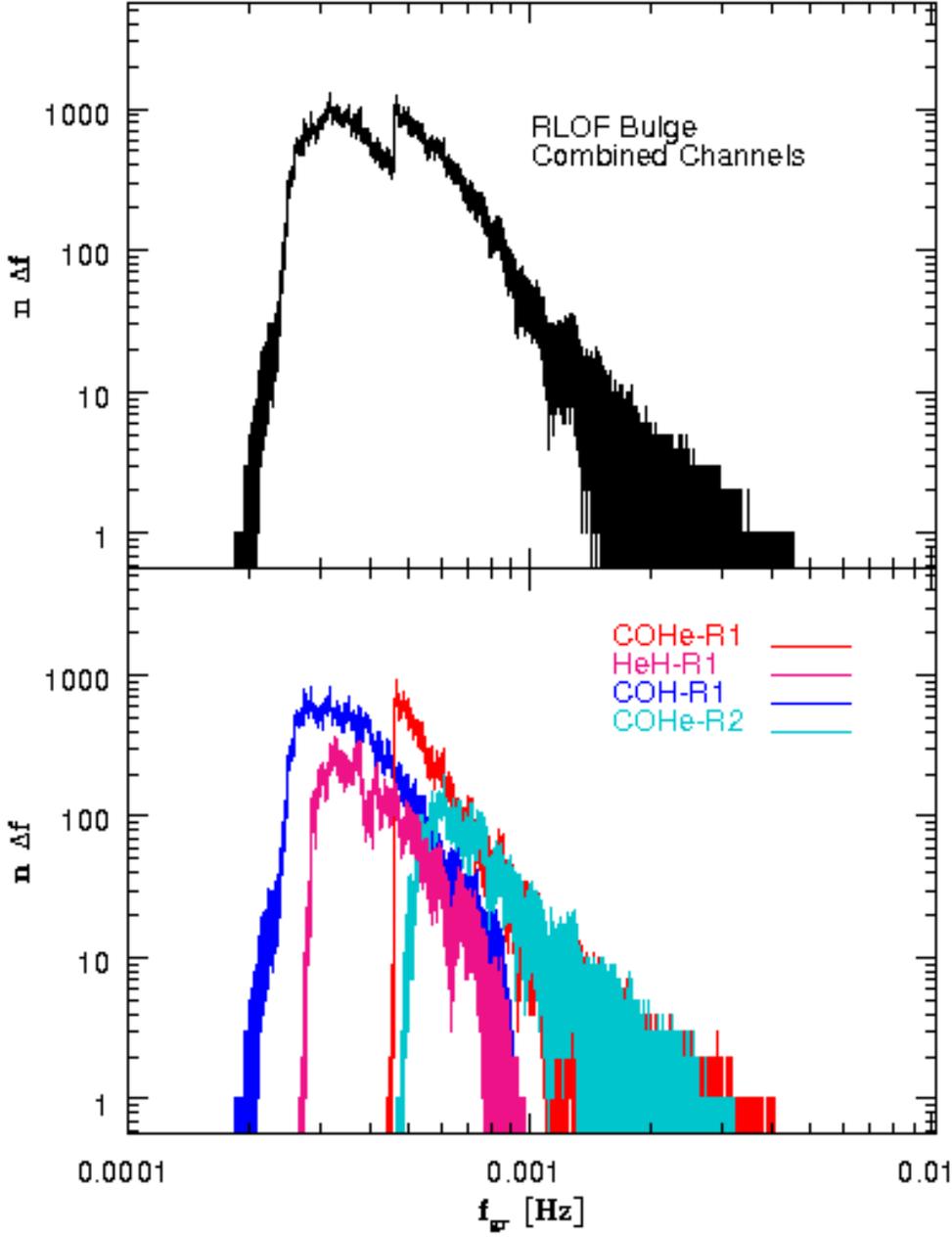}
\caption{Number density of the most prominent RLOF {\em LISA} double WD 
evolutionary channels for the Galactic bulge.  Note that COHe-R1 is 
descended from COHe-D1 in Figure~\ref{dndfbuldet}, in which the binary
undergoes two CE phases.} 
\label{dndfbulrlo}
\end{figure*}

In Figure~\ref{dndfdisdet}, Figure~\ref{dndfdisrlo},
Figure~\ref{dndfbuldet} and Figure~\ref{dndfbulrlo} we show in
(arbitrary) colour the number densities 
of the most common formation channels for the detached and RLOF
populations in the disc and bulge.  We do not include 
the less-populated formation channels in the plot or it would be very difficult 
to distinguish between channels.  We do however include some formation
channels involving brown dwarfs, since even though these binaries do
not significantly affect the overall GR signal (and we have ignored
them in the signal calculation), they are relatively
abundant, specifically in the bulge population.  

It is expected that the disc would host binaries which have evolved from a 
wider variety of formation channels than the bulge, since the age of
the disc binaries extends over a large range: $\sim$ few hundred Myr
to $9$ Gyr.  Note that high frequency systems 
only exist in the Galactic disc. There is a depletion of very high 
frequency (few minutes orbital period) RLOF systems in the bulge 
since the present bulge population only contains double white dwarfs with  
long-lived ($> 9$ Gyr) formation histories.  
All of the heavier (ONe+CO, CO+CO, CO+hybrid) binaries originating from more massive
progenitors which were born during star formation in the bulge have 
either long since merged, or in the
case of RLOF binaries, the stars has long since reached contact 
and are now exchanging mass in RLOF on slowly expanding
orbits.\footnote{We also note
  that in certain cases, particularly for bulge RLOF systems such as some AM
  CVn binaries, the donor has reached a lower mass limit of 0.01 
  M$_{\odot}$ at which point we no longer follow the evolution.}  
Once any system encounters stable RLOF the binary's orbital period 
increases as a consequence of conservation of angular momentum under
continued mass exchange, since the less-massive (larger) WD is losing 
matter to the more massive WD.  For the Galactic disc, some WD
binaries are still relatively young, and are in an earlier phase of 
stable RLOF and thus we `catch' these mass transferring systems at 
shorter orbital periods (GR frequencies above $\sim 4.5$ mHz; 
$P_{\rm orb} \lesssim 7-8$ min). 

{\it Typical detached evolution}:
Detached double CO-WDs are formed through a variety of evolutionary channels.
For the most prominent CO-CO channel (COCO-D1, see Table 1), one particular
example of evolution proceeds as follows: Two MS stars
($2.88$ and $2.45 \msun$) start out with a $P_{\rm orb}$ of $14.2$
days and an eccentricity of $0.54$ when the MW is $9180$ Myr old.
At $9602$ Myr, the more massive star (s1) begins to evolve off of the MS and
becomes a Hertzsprung Gap star.  Shortly thereafter s1 fills its Roche Lobe and 
begins to transfer matter to the companion (interaction between component stars 
aids in circularizing the orbit; $P_{\rm orb}$ $\sim 4.5$ days).  
RLOF is dynamically stable, and since s1 quickly becomes the
less-massive star during mass transfer $P_{\rm orb}$ increases
(to $\sim 143$ days), and RLOF stops at $9607$ Myr when s1 is a red 
giant ($0.44 \msun$; core mass $0.43 \msun$) 
and s2 ($3.67 \msun$) is still on the MS.  s1 continues 
to evolve and shortly thereafter becomes a (naked) helium  
star ($0.44 \msun$), as it has lost the remaining part of its depleted 
(by prior RLOF) envelope.  At $9731$ Myr, s2 has evolved off of the MS 
to become a Hertzsprung Gap star.   By $9777$ Myr, s2 has now evolved 
into an early AGB star ($3.62 \msun$), and the orbital period has decreased to $81$ 
days due to tidal interactions (expanding s2; s1 is still a helium star).  
s2 then fills its Roche Lobe, and since now the mass ratio is rather large ($M_{\rm don}/M_{\rm acc}
\approx 8$) the mass transfer is dynamically unstable, leading to a CE phase.
After the CE event the donor (s2) has become a Hertzsprung Gap helium star
($0.82 \msun$) and $P_{\rm orb}$ has decreased by more than $2$ orders of
magnitude to $3$ hours.  The orbit then starts to decay slightly due to angular
momentum losses associated with GR emission, and there is a brief RLOF phase
in which the Hertzsprung Gap helium star (s2) loses $\sim 0.1 \msun$, half of which is gained
by s1 (still a helium star, now $0.49 \msun$).  When the MW is $9779$ Myr old, 
s2 finally becomes a CO WD (primary WD).  s1 continues to burn helium and 
becomes a Hertzsprung Gap helium star at $9838$ Myr ($P_{\rm orb} = 2.7$ hours).
Within 20 Myr, s1 then becomes a low-mass CO WD (secondary WD) finishing its helium 
star evolution.  At $10$ Gyr (present) the system is found as a detached double CO WD with component
masses of $0.70$ \& $0.49 \msun$, and $P_{\rm orb} = 2.28$ hours ($f
= 0.24$ mHz).  The two WDs will be brought together in $\sim 300$ Myr due to GR
emission.  The system is expected to merge due to dynamically unstable mass transfer 
with a combined mass of $1.19 \msun$ (pre-merger orbital period of $\sim 1.3$ minutes).  
The total mass of the merged system does 
not exceed the canonical Chandrasekhar mass and so it is an unlikely Type Ia Supernova 
progenitor, though may result in the formation of a massive rapidly 
rotating WD \citep{SN04,YL05}.  

\begin{figure*}
\includegraphics[width=\textwidth]{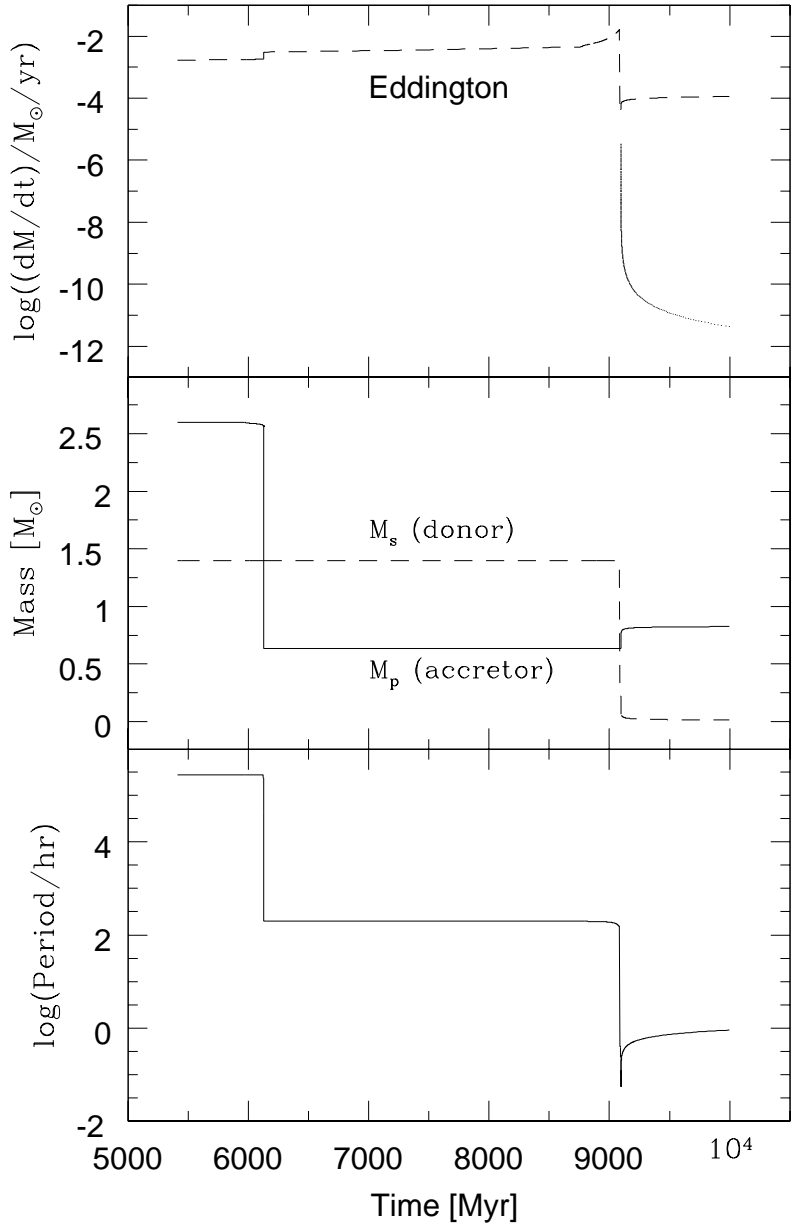}  
\caption{Characteristic evolution as described in \S\, 3.1 of a RLOF system 
with a CO WD accretor and a He WD donor, from the AM CVn formation channel COHe-R1 
starting at the ZAMS.  Top panel shows the mass transfer rate evolution (with 
the critical Eddington rate shown),  
middle panel shows the mass change of the two components, while the bottom
panel illustrates the orbital period evolution.  The two CE phases coincide  
with the prominent drops in orbital period when the donor (initial primary) is a late AGB 
star, and then when the donor (initial secondary) is a red giant.  Stable RLOF begins at the 
period minimum of $\sim 3$ min, when the Galaxy is $9.1$ Gyr old.  }
\label{rlof}   
\end{figure*}

{\it Typical RLOF evolution}:
In Figure~\ref{rlof} we show a representative evolution through RLOF for
one of the Galactic double WDs, a COHe-R1 channel system (see Table 2).  
The primordial binary consists of two low-mass MS stars ($2.60$ and $1.40 \msun$) 
on a wide ($P_{\rm orb}= 31.0$ year) and eccentric ($e=0.9$) orbit, born when 
the MW is $5410$ Myr old. At $5966$ Myr the more massive star (s1) begins to 
evolve off of the MS.  The orbit begins to shrink due to tidal interactions 
and by $6126$ Myr, s1 is an early AGB star, the orbit has circularized ($e=0$) and 
the period has decayed to $2.5$ years.  Within $1$ Myr s1 becomes a late AGB star.  
Magnetic braking (convective envelope of s1) dominates angular momentum loss  
and the orbit decays slightly.  s1 fills its Roche-Lobe ($6127$ Myr) and mass transfer is 
unstable, leading to a 
CE phase and a drastic decrease in orbital period (from $2.4$ years to 
$8.2$ days).  s1 then becomes a COWD 
(primary WD, $0.64 \msun$).  
At $8782$ Myr, s2 evolves off of the MS.  The orbit decays due to magnetic braking 
(convective s2), and at $9085$ Myr s2 initiates a CE phase resulting in another 
prominent period decrease ($6.1$ days to $30$ mins) and the loss of
its hydrogen-rich envelope.  s2 then becomes a He WD ($0.22 \msun$). 
GR causes the WDs to come into contact a few Myr later ($P_{\rm orb} \sim 3$ mins), and 
the secondary (s2) fills its Roche-Lobe.  The mass ratio is only $\sim 0.34$ thus   
RLOF is stable, and an AM CVn system is born.  Initially mass transfer is quite 
rapid (see Figure~\ref{rlof} top panel), but a slower mass transfer phase follows.  
The binary is found at $10$ Gyr with $P_{\rm orb} \sim 1$ 
hour ($\sim 0.6$ mHz) and component masses of $0.83$ and $\sim 0.01$ $\msun$ for the CO and 
He WDs, respectively.  Many of the
$25$ observed AM CVn binaries' parameters are uncertain, and
detection of these systems is biased against the typical, longer-period
systems with lower mass transfer rates.  However
this particular simulated binary's properties are somewhat 
similar to the very few long-period AM CVn binaries which have been
detected (e.g., SDSS J1552 \& CE 315), though the nature of the donors  
is still unknown \citep{And05,RNG07}).  

\subsection{Detached population characteristics}

\begin{figure*}
\includegraphics[width=\textwidth]{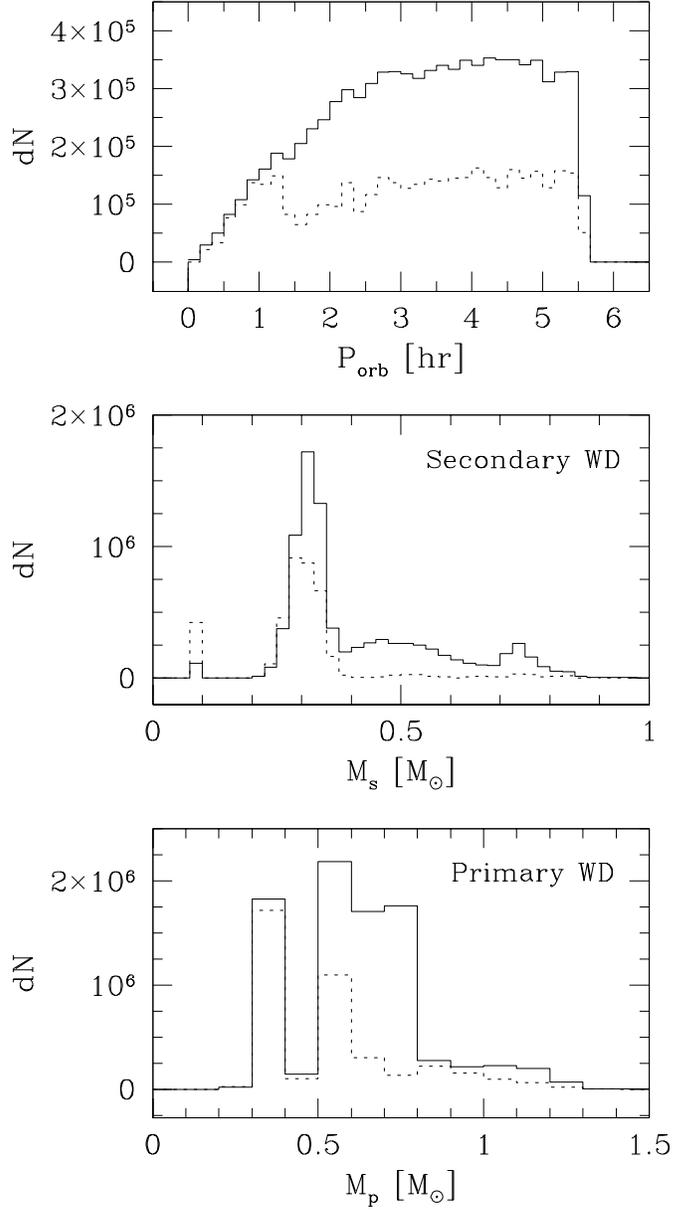}
\caption{Physical properties of the Galactic population of detached {\em LISA} double 
WDs.  The solid line shows the disc binaries, where as the dotted line shows the 
contribution from the bulge.  Bottom and middle panels show the distribution of primary 
and secondary mass (more massive and less massive WD, respectively), while the 
top panel shows the orbital period distribution.  Note the 
different bin size and scales used on the axes. }
\label{popd}
\end{figure*}

\begin{figure*}
\includegraphics[width=\textwidth]{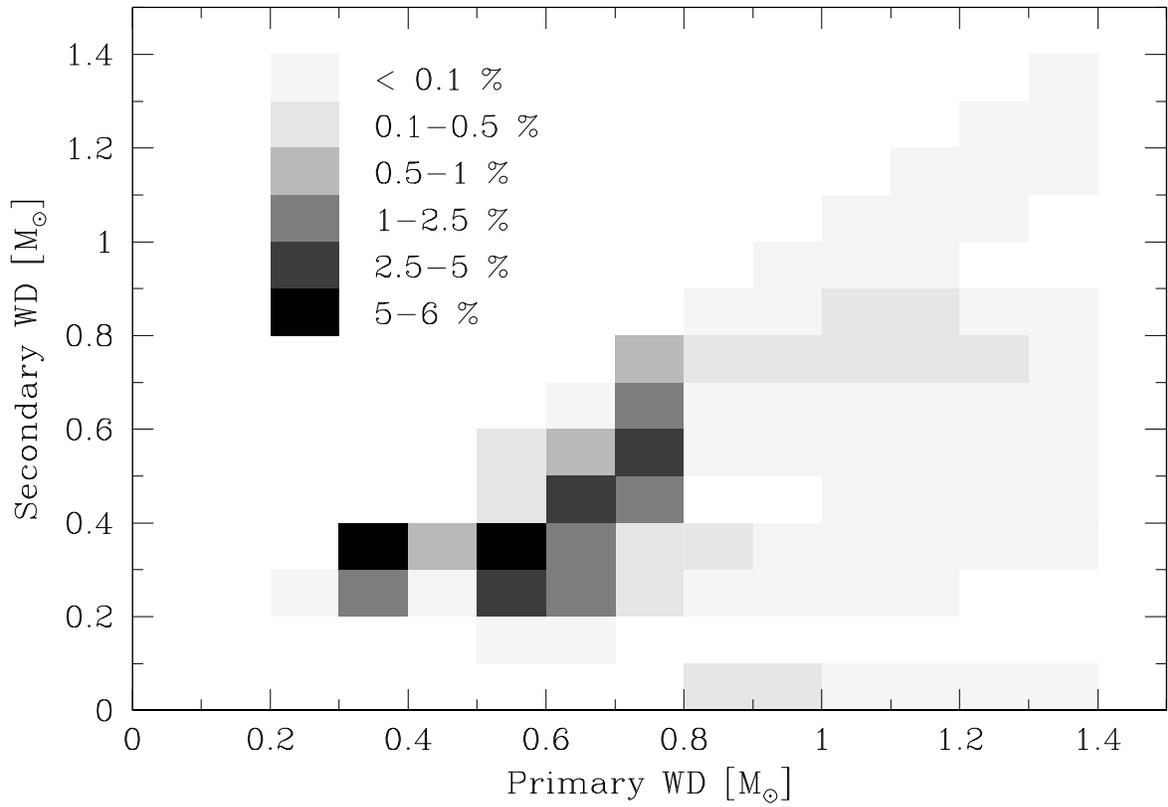}
\caption{Primary and secondary WD masses for Galactic disc 
  detached {\em LISA} binaries (corresponding to the solid line histogram in
  Figure~\ref{popd}).  The greyscale shows the relative contributions
  (percentage wise) relative to the Galactic disc {\em LISA} double WD population.}.  
\label{greydet}
\end{figure*}

Characteristic properties of the entire detached population are shown in
Figure~\ref{popd}.  Disc systems are found at all orbital periods
(top panel) within the {\em LISA} sensitivity range, with a drop in number 
at shorter periods due to the fact that once systems reach contact, 
double WDs either fall into the RLOF population or otherwise merge and drop
out of the {\em LISA} population altogether.  Secondary ($M_{\rm s}$,
less massive
WDs, middle panel) lie predominantly in four regions: $\sim 0.1 \msun$ (He WDs with CO
primaries), a prominent peak near $0.3 \msun$ (mostly He with some
hybrid WDs with mostly CO or He companions), a small clump near $0.5 \msun$
(CO secondaries with CO primaries) and a short peak at $\sim 0.7
\msun$ (CO or sometimes ONe secondaries with either CO or ONe
primaries).  The He secondaries with very low mass ($0.1 \msun$)
derive from the same formation channels as cataclysmic variables (similar systems 
have been discussed in \citet[][]{Pod08}), in
which a CO WD accretes from a low-mass main sequence star for several
Gyr (note the higher number of these systems in the Galactic bulge).
Eventually the donor becomes fairly exhausted of its hydrogen
supply, and meanwhile has built up a significant helium core.  The
donor eventually is depleted of so much mass that it reaches the
hydrogen burning limit (0.08 $\msun$), and thus becomes degenerate (in
this case, helium rich; a helium WD).  Many of the systems whose
secondary mass peaks near 0.3 $\msun$ are progenitors of AM CVn stars (COHe),
while most of the binaries involving more massive progenitors (CO, ONe) will likely merge once
they reach contact.  Primary WDs ($M_{\rm p}$, more massive WD; bottom
panel) display a peak at $0.3-0.4 \msun$,
though most primaries have masses $0.5 - 0.8 \msun$, and a smaller
number of systems are quite heavy ($0.8 - 1.4 \msun$).  The low mass
peak is comprised of He WD primaries, with He secondaries.  The
majority of the rest are CO WDs, with either He or CO secondary
companions.  In both the middle and bottom panels, we note that there are
relatively fewer heavy WDs in the bulge population.  This is because 
all of the more massive progenitors (e.g., CO-CO) have had a
sufficiently long enough time to evolve off of the main sequence,
encounter 1 or 2 CE phases, reach contact (GR is also more efficient at
bringing together two CO WDs than two He WDs) and have either merged out
of the population altogether or in some cases, have moved into the
RLOF population.  
A two-dimensional greyshade plot is presented in Figure~\ref{greydet},
where we show the relation between secondary and primary masses of
{\em LISA} binaries in the Galactic disc, presented in terms of 
relative total percentages of {\em LISA} double WDs.  

\subsection{RLOF population characteristics}

\begin{figure*}
\includegraphics[width=\textwidth]{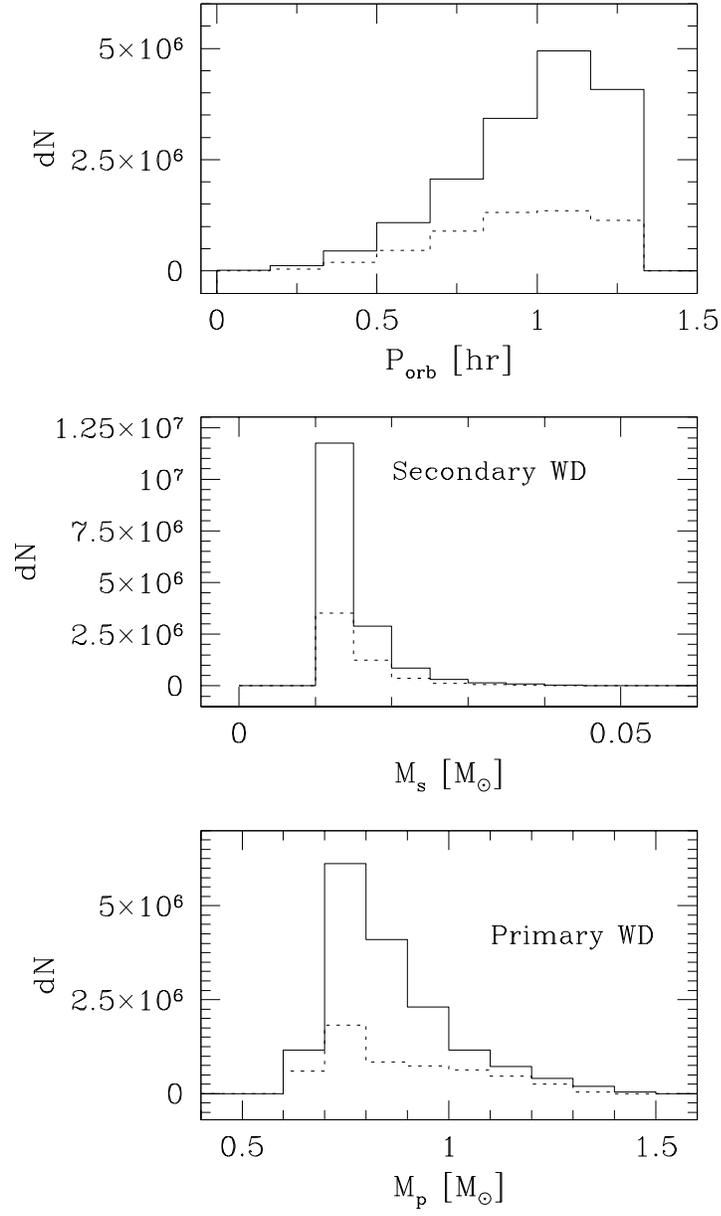}
\caption{Same as Figure~\ref{popd} but for RLOF systems.} 
\label{popr}
\end{figure*}

\begin{figure*}
\includegraphics[width=\textwidth]{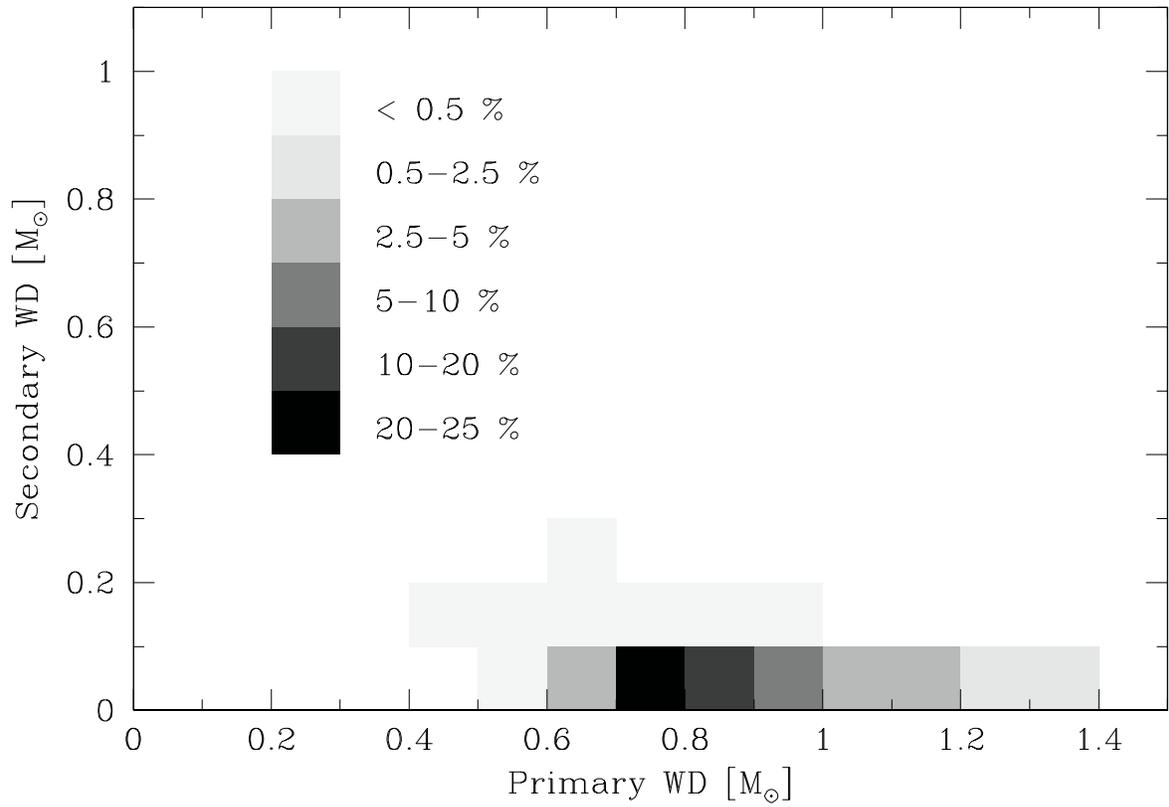}
\caption{Primary and secondary WD masses for Galactic disc 
  RLOF {\em LISA} binaries (corresponding to the solid line histogram in
  Figure~\ref{popr}).  The greyscale shows the relative contributions
  (percentage wise).  Note the different scaling on the y-axis as
  compared to Figure~\ref{greydet}}.  
\label{greyrlo}
\end{figure*}

Characteristic properties of the entire RLOF population are shown in
Figure~\ref{popr}. 
The RLOF systems with the maximum $P_{\rm orb}$ (lowest GR frequencies) appear at 
$\sim 75$ minutes (log($f_{\rm gr}$)$=-3.34$).    
The shape of the period distribution (top panel) is due to the fact that 
most systems spend the majority of their time at periods of $\sim 1$ hr during the  
dynamically stable RLOF phase.  As indicated previously, 99 \% of RLOF
systems are AM CVn (a WD accreting from a He or hybrid WD).  
For most WD binaries, at the onset of stable RLOF orbital periods 
are on the order of only a few minutes, and mass transfer rates are initially high 
(see Figure~\ref{rlof}, top panel).  The donor quickly becomes exhausted of 
most of its mass, the mass transfer rate decreases and the period continues  
to increase, though at a much slower rate.  In the middle panel we
show WD secondary masses, which is strongly peaked at $\sim 0.01
\msun$, with a slight drop-off to $ 0.045 \msun$.  
These WDs are comprised of mostly He WDs, some hybrid WDs and a small
number of CO WDs.  The `pile-up' of
WDs at $ 0.01 \msun$ stems from the fact that once a AM CVn WD-WD
progenitor reaches contact, initially the mass transfer rates are
high ($\sim 10^{-6} \msun$ yr$^{-1}$) and the donor is rapidly
exhausted of mass, thus it is rare to catch AM CVn secondaries which are relatively
massive ($> 0.02 \msun$).  In the bottom panel we show the
distribution of masses among the primary WDs, which are mostly CO WDs,
with a small contribution of ONe WDs (masses $> 1.3 \msun$).  
We show in Figure~\ref{greyrlo} a two-dimensional greyshade plot of
RLOF {\em LISA} binaries in the Galactic disc,  presented in terms of 
relative total percentages of {\em LISA} double WDs.

\section{{\em LISA} Signal}

During a one-year observation period, the width of a resolvable 
frequency bin is $\Delta f = 1~{\rm yr}^{-1} \approx 3 \times 
10^{-8}~{\rm Hz}$.  Below 1 mHz, there are hundreds to thousands of 
binaries per resolvable frequency bin, causing the signal to be 
confusion-limited.  In our simulation, below $\sim 0.45$ mHz the only contributing 
binaries to the {\em LISA} signal are the detached double white dwarfs.

\begin{figure*}
\includegraphics[width=\textwidth]{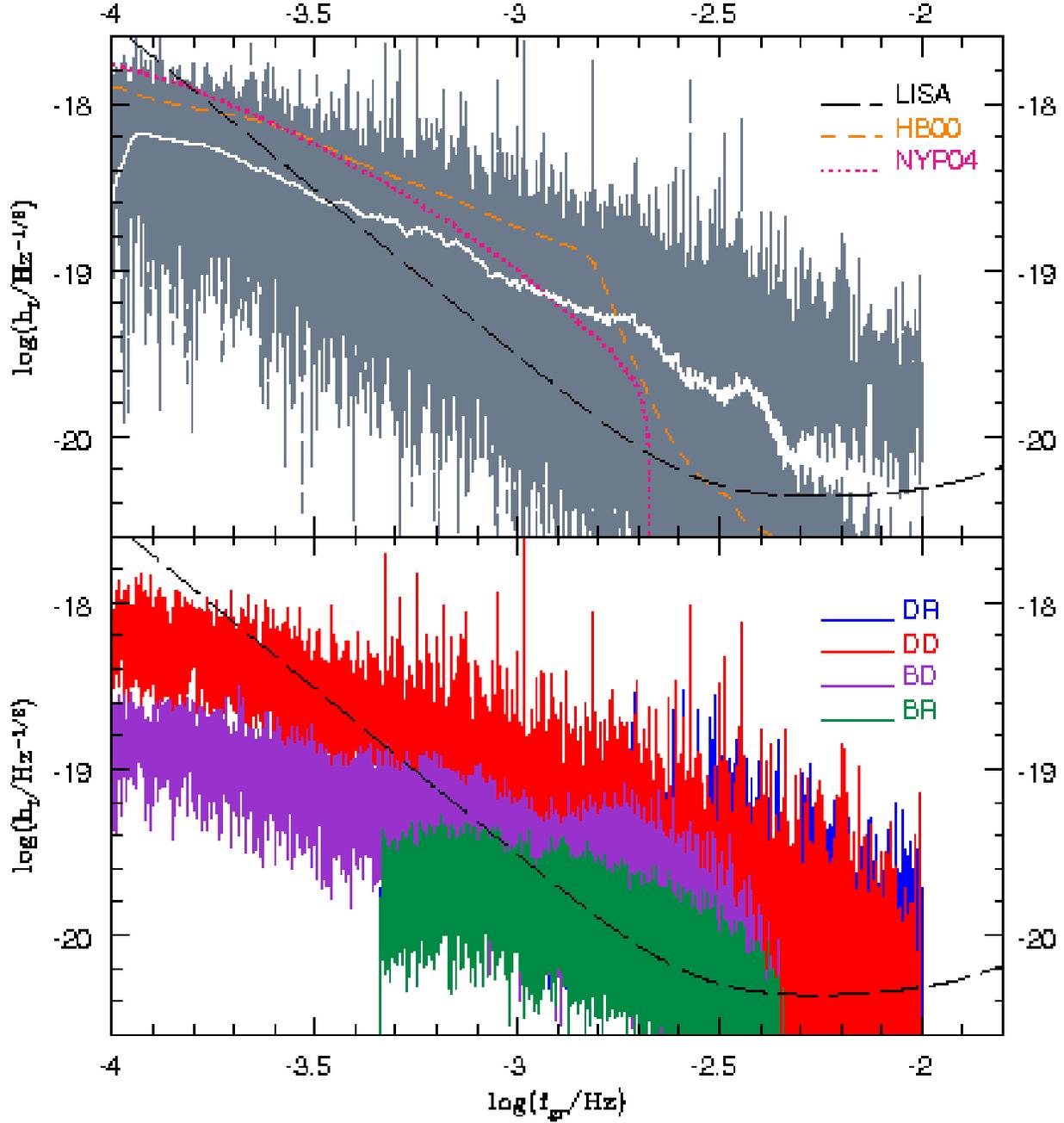}
\caption{Spectra (amplitude density) of the gravitational 
wave signal for {\em LISA} galactic WD binaries.  Bottom Panel:   
Disc detached (red), bulge detached (mauve), disc RLOF (blue) and
bulge RLOF (green).  Top Panel: The combined population is shown
with no smoothing (grey), and with a running median over 1000 bins (white).  Additionally, the
foreground estimates of \citet{NYP04} (dotted pink) and \citet{HB00} 
(dashed orange) are shown for comparison.  
Both Panels: Note that the signals have been truncated at frequencies 
above 0.01 Hz (see text).  The standard {\em LISA} sensitivity curve (black long-dash) for a
signal to noise ratio of 1 is also shown.}
\label{spectra}
\end{figure*}

The {\em LISA} spectra (amplitude densities, or spectral amplitudes
$h_{\rm f}$) for the Galactic double WDs were 
calculated using the sophisticated {\em LISA}
simulator of \citet{BDL04}, and are shown in Figure~\ref{spectra}.  
Close binaries which evolve to contact, such as 
precursors to the AM CVn population or progenitors of 
white dwarf mergers, 
will spend a very short time ($< 1$ Myr) at GR frequencies $> 0.01$ Hz 
($P_{\rm orb} \lesssim 3$ minutes, provided they reach these short periods), 
and this is reflected in our original 
{\tt StarTrack} calculations.  Thus, we artificially truncate the signal 
above 0.01 Hz since we simply 
do not have a large enough number of {\tt StarTrack} binaries above this 
frequency in order to accurately extrapolate them to the total number of 
Galactic systems here.  However, we expect the number of \textit{LISA} 
systems above 0.01 Hz to be $\lesssim 10^{3}$.  
In the bottom panel, we show the spectra from the four
Galactic sub-populations, while in the top panel we show the full spectra
for the combined population (grey) with a running median over 1000
frequency bins (white). 
The white curve represents our Galactic double WD
foreground (see \S\, 4.2).  We also show the foreground estimates of \citet{NYP04} 
(dotted pink) and \citet{HB00} (dashed orange) for comparison. 
The foreground of
\citet{NYP04} was artificially truncated beyond $\sim 2$ mHz, 
above which individual binaries were expected to be resolvable in that
work.  
It immediately comes to attention that the bulge binaries (green and
mauve; lower panel) do not contribute to the GR signal above
log($f_{\rm gr}$) $= -2.35$
($f_{\rm gr} = 4.5$ mHz) for reasons discussed in \S\, 3.1.  Also, the RLOF
binaries do not contribute to the signal at low
frequencies (below (log($f_{\rm gr}$)$=-3.34$, $f_{\rm gr}=0.45$ mHz) since these are binaries
which have been in a state of RLOF for a prolonged period of time, and
are on very slowly expanding orbits (largest orbital periods $\sim 75$
minutes).  Additionally, once the donor mass drops below 0.01 $\msun$, we stop the
calculation thus these very low-mass, low frequency systems do not contribute to the
calculated {\em LISA} signal.   

The RLOF disc signal (blue; lower panel) becomes comparable to the detached disc
signal (red; lower panel) at frequencies above $\sim$ log($f_{\rm gr}$)$=-2.5$ ($f_{\rm gr} \sim 3$ mHz).  
In our calculations, the AM CVn
binaries spend only a short period of time ($\sim 10^{1} - 10^{2}$ Myr) 
as detached double white dwarfs, where as they may spend $\sim$ Gyr in the 
RLOF phase.  This is likely a consequence of the common 
envelope prescription that we have employed, and most AM CVn systems
go through a double CE (COHe-R1 channel).  The \citet{Web84} CE
prescription with $\alpha \times \lambda = 1$ produces relatively
close post-CE white dwarfs, and so it does not take long for the stars
to be brought into contact via gravitational radiation once the detached 
double white dwarf has been formed.

In the upper panel, we show our combined signal (grey) and the median foreground
(white line) alongside the astrophysical foreground estimates of
\citet{HB00} and \citet{NYP04}, both of which include detached and AM 
CVn binaries.  
Note that our foreground is significantly
lower than the classic \citet{HB00} estimate at low frequencies, and
is more comparable to the curve of \citet{NYP04} in this regime (see
below).  Our foreground curve has been smoothed over 1000 bins, but
some `bumpiness' still exists at higher frequencies due to low number
statistics.  In contrast to some previous studies where the signal has been
truncated at higher frequencies, we show the curve over a
significant portion of 
the LISA bandwidth; up to 0.01 Hz. Since our original {\tt StarTrack} binary population was scarcely populated at
high frequencies, consequentially we could not generate smooth formation
channel PDFs (\S\, 2.2) for high-frequency systems.  

Our level of foreground signal is below that of \citet{NYP04} for GR frequencies $\lesssim
0.0013$ Hz.  This is in part due to the fact that at low frequencies, there 
is an overall lack of {\em LISA} binaries, since a large fraction of our {\em LISA}
binary population are 
RLOF systems and hence do not exist at the low end of the frequency 
spectrum.\footnote{We wish to point out that the number of binaries
  per resolvable frequency bin at 0.0001 Hz in Nelemans et al. (2001) is a factor of $\sim 2$
  higher than ours.}  The median chirp mass of our double white
dwarfs at frequencies below $\sim 0.0013$ Hz is $\sim 0.38$
M$_{\odot}$ (disc) and 0.31 M$_{\odot}$ (bulge).

Despite the large number of RLOF systems in part of the low frequency,
confusion-limited region of the Galactic gravitational wave spectrum 
(Figure~\ref{dndfdisrlo} and Figure~\ref{dndfbulrlo}),
the amplitude of the gravitational wave signal at low frequencies is 
dominated by the detached double WDs.  
Mass-transferring binaries at the lower
frequencies have undergone a significant amount of mass transfer, 
which leaves them with lower chirp masses than the detached binaries.   
As discussed previously, RLOF systems do not occur with orbital
periods greater than $\sim 2.5$ hours ($f_{\rm gr} \lesssim 0.45$ mHz).  
Consequently the gravitational wave amplitude of the
RLOF systems (both from the disc and from the bulge) does not 
contribute very strongly to the overall foreground noise in the 
low-frequency regime.  The GR amplitude from the bulge is smaller 
than that of the disc for both (corresponding) detached and RLOF 
components of disc, primarily due to the fact that the stellar mass 
of the bulge is $\sim 1/3$ the mass of the disc, and contains 
fewer {\em LISA} double WDs.  

\subsection{Transition Frequency}

The `transition frequency' is said to correspond to the GR frequency
at which the gravitational wave spectrum transitions from confusion-limited to 
individually resolvable sources.  The exact conditions which are
necessary in order for this to occur (one possibility would be when
the average number of binaries per frequency bin
drops below 1) are not completely clear.    
The addition of RLOF systems to the 
detached population will increase the frequency at which the average number of 
binaries per bin drops below 1, simply by adding more sources to 
the population.   This is particularly so since RLOF binaries are able
maintain short orbital periods (for longer then their detached
counterparts) once contact has been established.  Furthermore, as the transition frequency  
increases, the spread of the signal due to the motion of {\em LISA} also 
increases, thus the signals will continue to overlap at higher 
frequencies than might be inferred from a simple analysis of the 
number of binaries per resolvable frequency bin.   
For a simple detached system in a circular orbit whose orbital period
is evolving solely due to the emission of GR,
seven parameters are needed to fully characterize the signal.  Basic
estimates from information theory describe how the limited amount of
information in the {\em LISA} data stream can be mapped into an equivalent
amount of information about detected binary systems.  A common rule
of thumb is the ``three-bin rule'': at least three (otherwise unpopulated) frequency bins of
{\em LISA} data are needed to resolve a seven-parameter binary source 
\citep[see][and references therein for a discussion]{TRC06}.  One
can reasonably expect that a mass-transferring system will require
more parameters to characterize the GR signal, and
consequently more frequency bins to individually resolve the source.
Therefore, the average number of binaries per frequency bin is not a 
particularly good indicator of how the transition frequency changes with the 
addition of the RLOF population.

For the sake of comparison with previous studies, we have estimated the 
transition frequency using the 3-bin rule method.  The point at which
{\em LISA} binaries {\em start} to occupy one source per three
frequency bins is found to be 1.5 mHz for the detached {\em LISA} binaries
only, which is comparable to \citet{NYP01} and \citet{Nel03b}, who found a transition 
frequency of $1.5 -2$ mHz see also \citep[][for a discussion
of the inclusion of AM CVn binaries in the confusion limit
calculation]{NYP04}.  However we find that when we consider the
entire {\em LISA} population of double WDs, the transition frequency
is increased to 2.4 mHz.  We would like to
however point out that these frequencies do not represent the frequencies
above which one expects to resolve all {\em LISA} binaries; there are
still a number of populated bins at higher frequencies.  

The 3-bin rule method of estimating the
frequency at which binaries start to become resolved is not a robust 
method, since it is expected that at these higher frequencies, the 
frequency evolution due to mass transferring binaries can contribute 
a significant additional spreading of the signal.
Further, this method does not take into account important
criteria such as Doppler, phase and amplitude modulation, 
and one should consider alternative (more technical) methods
in computing the threshold frequency region \citep[e.g., Chapter 2 of][]{phd}.  
In the following subsection, we
discuss an approach other than the 3-bin rule of thumb to identify the 
population of resolved sources.  Instead of using a general estimator like the transition
frequency, it is possible to use a predicted signal-to-noise ratio for
any particular system to determine whether it is potentially
resolvable or not \citep[see also][]{TRC06}.  

\subsection{Noise}

For {\em LISA} observations of length $T_{obs}$ the sensitivity to a
particular binary
source can be expressed by constructing a simple estimator of the
signal-to-noise ratio (SNR).  For circularized compact binaries, long
observations will narrow the bandwidth of the source, and the binary
will appear in the Fourier spectrum as a narrow spectral feature with
root spectral density $h_{b}(f)$, given in terms of the strain
amplitude $h_{o}$ as \citep{LHH00}
\begin{equation}
	h_{b}(f) = h_{o}\,\sqrt{T_{obs}}\ .
	\label{binarySpectralAmplitude}
\end{equation}

\begin{figure*}
\includegraphics[width=\textwidth]{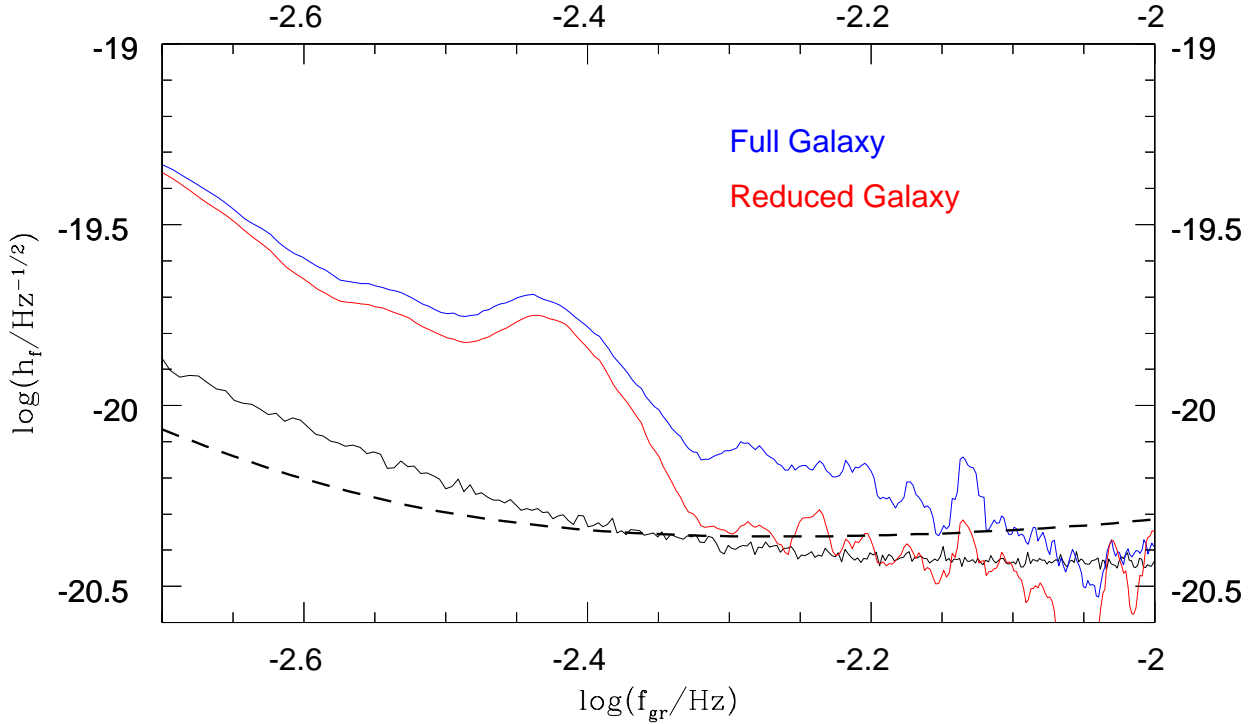}
\vspace{-2cm}
\caption{Median spectra shown in Figure 10 (smoothed): the median Full Galaxy 
alongside the median Reduced Galaxy (red; full median signal with resolved sources removed), 
the standard {\em LISA}
sensitivity curve (dashed line; SNR=1, also shown in Figure 10) and the simulated 
{\em LISA} Michelson noise curve (\S\ 2.3).  As in Figure 10, the 
spectra have been truncated above 0.01 Hz.  Both the Full Galaxy {\em LISA} double
white dwarf foreground (blue line) and the Reduced Galaxy foreground curve
(red) fall below the Michelson noise at roughly $\sim 8$ mHz, where
the curves are relatively noisy.  When one considers
the median foreground spectra against the standard sensitivity curve, the Full
Galaxy foreground drops below the sensitivity curve at $\sim 7.5$ mHz,
while the Reduced Galaxy foreground drops below the sensitivity curve
at $\sim 6$ mHz.  The
(black) noise curves differ in sensitivity level and shape since the underlying
noise levels used to generate each noise curve are different.  In
our study we have used the Michelson noise curve to represent the LISA
instrumental noise; the standard sensitivity curve is shown for comparison.}
\label{curves}
\end{figure*}

{\em LISA} will detect binaries in the presence of competing sources of
noise.  There will be two components to the noise in {\em LISA} analysis:
instrumental noise in the detector itself, and irreducible
astrophysical noise from the total population of Galactic close
binaries.  The instrumental component of the noise is given by the
shape of the {\em LISA} sensitivity curve \citep{LHH00}.  In this
work, the Galactic foreground noise is computed from the 
synthesized Galaxy itself (i.e., with {\tt StarTrack}).  
The gravitational wave strength of every binary in the synthesized Galaxy is estimated
from its parameters, and a running median of the total signal is computed as a function
of frequency.  Since the total number of individually resolved binaries is expected to be
small compared to the total population, the running median should be a good estimate of
the residual astrophysical noise foreground that would result from a fully realized
science analysis procedure.  The foreground resulting from this procedure has been shown
in Figure~\ref{spectra} (top panel).  Additionally, we have calculated
the median foreground signal which is expected to arise from the
unresolvable sources alone (the `Reduced Galaxy'), which is likely a
more realistic estimate of the true Galactic foreground.  In
Figure~\ref{curves}, we show the Galactic foreground 
alongside the Galactic foreground signal minus the signal arising from the
sources which were determined to be resolvable using our
signal-to-noise estimate technique, assuming a SNR ratio $> 5$,
(see \S\, 4.3).  Alongside our 
Galactic foregrounds we show for comparison 
the standard {\em LISA} sensitivity curve as well as the average Michelson
noise of {\em LISA}.  The latter 
curve (see \S\, 2.3) is a more appropriate comparison of the {\em LISA}
instrumental noise with our calculated gravitational wave signal from
the Galactic binaries, as both curves (our signal and the 
Michelson noise) account for directional and 
frequency dependencies of the wave as measured by {\em LISA} \citep{RCP04}.
The standard {\em LISA} sensitivity curve 
\citep{SCG} is often presented in the literature, and is an
approximation of the spectral amplitude h$_{f}$ scaled from the
barycentred sensitivity amplitude h$_{o}$ from the \emph{Online Sensitivity Curve Generator}.  

The ultimate level of the confusion foreground (and as a result, the
ultimate SNR of any given source) will depend strongly on the actual
data analysis algorithm used to identify and subtract sources.  To
date no fully realized implementation of such a procedure has been
developed.  The computed foreground shown here should be a good
estimate to the output of a fully developed analysis procedure.

\subsection{Resolvable Sources}

Though millions of double WDs are expected to be detected with {\em
  LISA}, a much smaller number of these systems are predicted to be 
resolved, depending upon the physical properties of each binary 
(see also \citet{SVN05} for a discussion).  Due to the small 
separations of these compact binaries, \citet{CFS04} predicted that 
nearly $1/3$ of the resolved {\em LISA} binaries will be eclipsing 
in the optical regime and thus will offer an opportunity for these 
systems to be studied with follow-up observations, allowing for 
the determination of WD physical parameters (i.e., radii).  It has 
been noted by \citet{Nel09} however that the intrinsic brightness 
of white dwarfs was overestimated in \citet{CFS04}, thus the 
expected number of 
{\em LISA} double white dwarfs with electro-magnetic counterparts 
should be lower.  \citet{Nel09} predict a smaller, though still 
relatively optimistic number of {\em LISA} systems which may have 
optical and/or near infrared counterparts; possibly as many as $\sim
2000$ sources (see their \S\, 3.3).

The total effective noise in the detector (instrumental $+$ foreground)
is given by the spectral amplitude $h_{f}(f)$, and the SNR is
estimated as \citep{LHH00}
\begin{equation}
	SNR \simeq \frac{h_{b}(f)}{h_{f}(f)} =
	\frac{h_{o}\sqrt{T_{obs}}}{h_{f}(f)}\ .
	\label{simpleSNR}
\end{equation}

A binary is deemed ``resolvable'' (distinguishable from the confusion
limited Galactic foreground) if its SNR is greater than some detection
threshold (typical SNR $\geq 5$).  
Additionally, resolved sources in the population are sorted into
\emph{monochromatic sources} and \emph{chirping sources}. 
Monochromatic sources do not evolve appreciably over the {\em LISA}
observing time, $T_{obs}$.  A binary is considered to be chirping if
its frequency changes by more than the {\em LISA} frequency resolution,
$\Delta f \geq 1/T_{obs}$.  This criteria is a conservative one, as
the expectation is {\em LISA} will be able to detect frequency changes which
are smaller than the frequency resolution $\Delta f$ \citep{TS02}.
The detection of binaries with appreciable evolution in $f$ (or `chirp')
is useful, since this frequency evolution allows for the determination of
${\cal M}$, from which the distance to the source can be calculated
\citep[see e.g.,][]{NYP01}.   

The estimate of the SNR in Eq.\,(\ref{simpleSNR}) will not be precise in the 
regime where orbital
periods change on timescales short compared to the observation time,
or where the shape of the instrumental noise changes appreciably as a
function of frequency.  However, this simple estimator is a good tool
to use as a first cut in a search over a large population of sources
where a calculation of the SNR for each binary would be
computationally prohibitive.  
For the purposes of this paper, this
estimator should be perfectly adequate since in large part the
binaries of interest are not chirping across a significant portion of
the {\em LISA} band during $T_{obs}$.  Once a set of Galactic binaries has
been identified using the simple estimator in Eq.\ (\ref{simpleSNR}),
it is computationally plausible to consider a more robust estimation
of the SNR for each system which can then be constructed from the data
analysis technique of choice.  For example, with chirping binaries a
good estimate of the SNR can be motivated from matched filtering
techniques by integrating over a binary's spectral energy
distribution, $dE/df$ \citep{FH98}. 

Using this simple SNR estimator to resolve {\em LISA} binaries, 
we find that a total of $\sim 11300$ double white dwarfs will be resolvable
in our Galactic sample: $\sim 600$ chirping and $\sim 10700$
monochromatic.  This is inferred 
under an optimistic assumption that all binaries with a SNR $>5$ have a chance to be 
resolved.  Within the resolvable population, we find $\sim 4000$ AM CVn
binaries and $\sim 500$ possible Double Degenerate Scenario - DDS;
\citet{IT84,Web84} - SN Ia
progenitors (detached COCO WD binaries which will
merge within a Hubble time, whose total mass exceeds the Chandrasekhar
mass limit).  Out of these possible SN Ia progenitors, the majority are formed through 
the COCO-D4 channel, which involves the most massive primaries on the ZAMS.  
As for the AM CVn resolvable
binaries, 7 \% are chirping and arise mostly from COHe-R1 and COHyb-R1
channels, in near-equal numbers.  Both
formation channels involve two CE events and thus if these systems
will be resolved with {\em LISA}, they will serve as useful objects for
understanding close binary evolution.  Additionally, a number of
progenitor AM CVn binaries are resolvable ($\sim 3000$; detached
channels), but only $\sim 100$ of these are chirping.

If we relax our criteria and assume that any binary with a
SNR $> 1$ is potentially resolvable, we find $\sim 43000$ resolvable double
WDs ($\sim 1000$ chirping, $\sim 42000$ monochromatic).  Additionally,
we find $\sim 2800$ {\em LISA} binaries with SNR $> 10$, $\sim 300$ of which are
chirping sources.  The binaries with large SNR ($\gtrsim 10$)
typically all have orbital periods $< 2000$ s ($f_{\rm gr} > 0.001$
Hz) and have either just recently encountered RLOF, thus they still
have relatively large chirp masses, or they are close to reaching
contact.  

In Table 5 we show a breakdown of the $11300$ (2800) potentially resolvable 
{\em LISA} double WDs which have SNR $> 5$ (SNR $> 10$), with percentages
relative to the resolvable population in question.    
The properties of resolved systems reflect some of the selection effects 
that are expected to arise from using {\em LISA} to identify  
double WDs. Systems with short orbital periods (and consequently high 
gravitational wave frequencies) are favored both because the amplitude of 
the signal scales as $f^{2/3}$ and because the number of binaries per bin 
drops with increasing frequency. Furthermore, systems with large chirp 
masses are favored simply because the amplitude of their  
GR scales as ${\cal M}^{5/3}$. Thus, we see that the resolvable 
population is biased toward high mass, high frequency 
(short period) systems.  
Realistically speaking, it is still unclear as to which method(s) of detecting resolvable 
binaries with {\em LISA} will be most successful, and so we have
discussed the likelihood of potentially 
resolving individual binaries using a few different methods.    

{\em Detached resolved sources.}  A large number of the 
resolved sources are of the HeHe-D1 channel, where two 
initially low-mass ($\sim 1 \msun$) evolved stars lose their envelopes
during mass transfer phases.  HeHe-D1 are 
not only the most numerous channel of the detached systems 
(Table 1 and Table 3), 
but also extend to high GR frequencies.  Lumping all 
of the COCO-D formation channels
together, COCO detached binaries make up $\sim 13-18$ \% of the resolved
population, and potential DDS progenitors make up 5\% of the total 
resolved population.
Though HeCO-D1 double WDs are more numerous compared to COHe-D1 binaries, 
fewer of them are resolved.  We note that 
COHe-D1 progenitors involve two CE phases (HeCO-D1 only one), and 
overall occupy closer orbits thus contributing to higher GR
frequencies.  

{\em RLOF resolved sources.}  For RLOF systems, the majority are of AM
CVn-type.  The most common RLOF channel is   
COHe-R1, followed by COCO-R1.  COCO-R1 binaries are not common, but they lie on
very close orbits, consisting of a massive ($\sim 1$ M$_{\odot}$) CO
WD accreting matter from a low-mass ($< 0.2$ M$_{\odot}$) CO WD.  
The resolved COHe-R1 AM CVn systems account for most of the total
number of resolved AM CVn binaries.   
Since the formation of these systems involves 
two CE events, future {\em LISA} observations may put some very useful constraints on this 
rather uncertain evolutionary phase which is crucial to the formation
of close binaries.  However, when comparing relative formation scenarios of
AM CVn binaries, we must keep in mind that the
precise behaviour of the stars upon reaching contact - and the
exact physical conditions of the donor at contact - are still
unclear, as are the physical characteristics of the AM CVn 
progenitors themselves.  Further observations and detailed modeling will
be an important step toward understanding the donors in AM CVn type systems \citep{DTW07}. 
For example, deviations from the equilibrium evolution of AM CVn systems, e.g., brought on upon by 
changes in the mass transfer rate following a nova explosion, result in 
a measurable frequency evolution which can greatly differ from that 
which is expected for the long-term, equilibrium evolution case.  This was already shown by 
\citet{SN09}; if short-term changes in the evolution of AM CVn systems are 
neglected, it can lead to erroneous estimations of some of the binary parameters.

\section{Comparison with Previous Studies}  

The number of binaries important in context of the {\em LISA} GR
signal was estimated by \citet{HBW90}  
and in their work they predicted $3 \times 10^7$ Galactic close double
WDs (orbital periods 
$\lesssim 1$ day) based upon a surface density 
star formation rate.  
However that estimate seemed to over-predict 
the observed local density of double 
WDs available at that time.  They concluded that even if 
they decreased the estimated space density by a factor of 10 
(decreasing the total number of Galactic double WDs), close WD
binaries would continue to dominate the {\em LISA} 
GR signal at low frequencies \citep[][\S\, 1]{HBW90}.  
Population synthesis studies \citep[e.g.,][]{NYPV01} which 
incorporate current estimates of Galactic 
star formation rates and SNe rates result in a higher 
number of Galactic double WDs 
($\gtrsim 21 \times 10^7$)\footnote{We find $\sim 5 \times 10^{8}$
  double WDs in the Milky Way.}, albeit the majority of these 
would have orbital periods too large to 
be detectable with {\em LISA}.  Still, we note here 
that current estimates of the number of 
double WDs with GR frequencies detectable with 
{\em LISA} are a factor of
$\sim 10$ higher than those used in 1990.

To populate the Galaxy (the disc only), \citet{HBW90} used the density distribution
\begin{equation}
\rho  = \rho_0 e^{-R/R_0} e^{-\left|z\right|/z_0},
\end{equation}
where $\rho_0$ is the central density, $R$ and $z$ are galactocentric 
cylindrical coordinates, and $R_0$ and $z_0$ are the radial scale 
length and vertical scale height of the Galactic disc, respectively.  
\citet{HBW90} used $z_0=90$ pc in their study, and while though a reasonable assumption for the scale 
height of the thin disc, likely led to their (too) low estimate of double WDs.  More recently, 
it has been noted \citep[e.g.,][]{BH06} that the scale height for the binary population 
may be much higher ($z_0 \sim 200$ pc) than that used in \citet{HBW90}.

The \citet{HB00} curve is a modified result based on \citet{HBW90},
who in calculating the GR amplitude in 1990, used too low of a space-density
estimate for double WDs.  
In the original study of \citet{HBW90} which neglected RLOF systems, 
Galactic double WDs begin to dominate the GR amplitude at $\sim 0.1$ mHz,  
and continue to dominate until $\sim 1.6$ mHz, where the average number of sources 
per resolvable frequency bin drops below 1 (similar results for the transition frequency 
were found in their study that included RLOF systems; \citet{HB00}).
The analytical study of \citet{HBW90} was not directly comparable to more recent
studies of the {\em LISA} foreground which utilized more sophisticated
techniques.  
\citet{TRC06} performed a re-calculation of the analytical \citet{HBW90} 
GR foreground for Galactic binaries in order to compare the results
directly to those of 
\citet{NYP01} \citep[see][their Figure 10]{TRC06}.  Little was known
about the true space density of double white dwarfs at the time of
the \citet{HBW90} study, and since
\citet{TRC06} was based on these results which
underestimated the space density (and hence total number) of Galactic
white dwarf binaries, the \citet{TRC06} study had a 
$90$\% decrease in source density as compared to the study of
\citet{NYP01}.  
This reduction in number of systems resulted in a decrease in the
foreground GR amplitude.  
Until more recent observations of double white dwarfs
were available to better constrain the space density
\citep[][]{Warner95,RNG07}, 
the error went undetected for some time, as 
this decrease was serendipitously counteracted by an increase in
amplitude imposed by the (too) large double WD chirp masses used in
the \citet{HBW90} and \citet{TRC06} studies.  In short, the studies of \citet{TRC06} (recalculated
\citet{HBW90}) and \citet{NYP01} have comparable foreground
levels, albeit for different physical reasons.  An increased number of observations 
of WD binaries since the time of the \citet{HBW90} study \citep[i.e., SPY project] 
[]{Nap04,And05} have confirmed that the (lower) WD masses used by
\citet{NYP01} coincide more closely with the masses of WDs in
binaries, where the masses used in \citet{HBW90} do not \citep[see also][for
further discussion]{Nel03b,TRC06}.

Based on \citet{EIS87}, \citet{NYP01} used the criterion that the transition frequency occurs 
when the average number of binaries per bin drops below 1, and they found the 
transition at $\sim 1.6$ mHz. They also found that $12124$ detached 
double WDs will be resolved with {\em LISA}, assuming all systems above the {\em LISA} 
sensitivity limit (a signal-to-noise of 1) are detectable.  
In their study which included AM CVn systems, 
\citet{NYP04} found a transition frequency closer to $\sim 2$ mHz, and
that a total of $\sim 22000$ Galactic double WDs (half detached and
half RLOF) would be resolvable.  For a signal-to-noise ratio of 1, 
we find that $\sim 43000$ double white dwarfs with GR 
frequencies $< 0.01$ Hz are potentially
resolvable with {\em LISA}.  However, in order to perform a reasonable comparison with
\citet{NYP01}, we need to consider only those binaries whose GR
frequencies are above $\sim 2$ mHz (and below 0.01 Hz; see \S\, 4).  
In this case, we find that $\sim 25000$ double white dwarfs may be resolvable with LISA. 

In previous studies, it was assumed that the level of the Galactic
foreground will not contribute above the frequency at which the
average number of sources per bin drops below 1 (coined
the `transition frequency').  Since
those studies do not calculate the GR signal beyond the transition
frequency, above $\sim 2$ mHz their results cannot be compared with
ours.  
We find that the frequency at which the WD foreground drops below 
the instrumental noise level ($\sim 6-8$ mHz) is above
the transition frequency which was found in previous studies.  
In contrast to previous work, it is expected that at higher 
frequencies ($\gtrsim 3$ mHz) the RLOF systems may 
contribute more significantly to the {\em LISA} signal than detached systems.  This
has implications for the `transition frequency' between
unresolved and resolved systems: the transition frequency is shifted
to higher frequencies with the addition of RLOF systems. 

Since we do not follow the evolution of non-degenerate 
stars in our {\em LISA} calculations, the binaries involving `pre'-hydrogen
WD donors (i.e., CVs) are not shown on the figures, though we note they would 
make some contribution to the {\em LISA} data stream.  
We do not include the `brown 
dwarfs' in our LISA signal populations, though we note that binaries 
involving these brown dwarfs will not contribute 
significantly to the Galactic gravitational wave background as these 
binaries have very low chirp masses.  Since the formation of such
objects takes on average several Gyr, it is however important to
include them when considering very old stellar populations such as the
Galactic halo \citep{Rui09}, where these systems comprise over $70$ \%
of the entire remnant binary population. In that work, it was found
that the {\em LISA} GR signal from halo white dwarfs will be at least an
order of magnitude below the signal arising from Galactic
(disc$+$bulge) white dwarf binaries.

\section{Summary and Conclusions}

 We have used the population synthesis binary
  evolution code {\tt
  StarTrack} and the detailed {\em LISA} simulator of \citet{BDL04} to
  calculate the gravitational wave foreground arising
  from close white dwarf binaries in the Milky Way disc and bulge.
  Our population synthesis includes both detached and
  mass-transferring binaries, and it is found that the
  mass-transferring (RLOF) binaries begin to contribute significantly to the foreground noise
  at high GR frequencies ($\gtrsim 3$ mHz, Figure 10 bottom panel).  

Our results are in general agreement with the analytical study of 
\citet{HB00} and the population synthesis study of \citet{NYP04} 
for the low frequency regime: the {\em LISA} signal is dominated by detached 
binaries. However, in contrast to previous studies, it is found that,
despite the fact that the `transition frequency' is found to be $\sim
2.4$ mHz using the 3-bin rule of thumb, at frequencies above $\sim 3$ mHz the presence of
RLOF binaries will likely make it difficult to resolve individual
sources, and we find that the {\em Galactic foreground dominates the LISA noise up to
at least $\sim 6$ mHz}.   
This is because the RLOF binaries add to the total number of {\em LISA} 
systems at high frequencies, and at high frequencies the RLOF 
systems chirp backward and spread out over more frequency bins,  
which offers a new set of challenges for
{\em LISA} data analysts. 
The details of the true orbital period evolution due to 
mass transfer and possible tidal effects \citep[e.g.,][]{DB03,Bil06} 
are not well understood, and this uncertainty may impact the ability of 
future data analysis schemes to successfully resolve these systems.  
It is starting to become clear that such effects should be taken into
account when studying the population of compact binaries
\citep{RPA07}.
One new and interesting result of our study is the possibility 
that RLOF systems which have recently entered the mass transfer phase
(small periods, higher chirp masses) can add to the confusion foreground at higher
frequencies than were previously predicted for the double white dwarf
population.

It should be repeated here that in this study, we employed one model prescription of CE 
evolution.  This important phase of binary evolution may have a large impact on a synthesized 
stellar population, and various model prescriptions of CE may lead to a different number of 
Galactic binaries and binary physical properties, which in turn affect the shape and strength 
of the {\em LISA} GR signal.  The number of close ({\em LISA}) binaries could potentially be 
decreased if the CE efficiency is low (more mergers) or if angular momentum balance 
rather than energy balance is assumed during the CE phase.  It was
shown in \citet[][]{RBH06} that
when an angular momentum balance prescription is adopted \citep{NT05},
post-CE orbital separations are overall larger, and fewer binaries
will enter a CV-like RLOF phase between a white dwarf and a main
sequence star, which is a crucial step for some of the older (bulge
COHe-R2) {\em LISA} binaries. 

We have shown that there is a large population of potentially resolvable
binaries in the Galaxy.  
The approach of estimating resolvable systems 
described in \S\, 4.3 (`SNR $\geq$ 5'), relies on assuming that
signals arising from individual 
binaries are separable, i.e., can be demodulated using some sort of matched 
filtering technique \citep[see][and references their-in]{Set02}.
If detected, once their properties are
measured these systems 
may aid to solve several remaining problems in binary evolution theory.
Especially interesting would be the detection of any double white dwarf
binary with a mass close to or over the Chandrasekhar mass (potential SN
Ia progenitor or possible pre-NS accretion induced collapse object), 
or the measurement of the orbital periods of systems which have just
emerged from a common envelope (the phase that determines the formation of
most compact object binaries).  However, we reiterate that the
detection of resolvable binaries will depend on the development of
sensitive detection schemes and data analysis techniques.

\acknowledgements 
We would like to thank Gijs Nelemans for 
very useful discussion on this project, which greatly improved this work, 
and for providing data for \citet{NYP01} and \citet{NYP04}.    
KB, MB and SLL
acknowledge the hospitality of the Aspen
Center for Physics.  MB and SLL were supported at the Aspen Center by
NASA Award Number NNG05G106G. MB is also supported by NASA APRA grant
Number NNG04GD52G. KB and AJR acknowledge support through KBN Grants 1
P03D 022 28 and PBZ-KBN-054/P03/2001, and the hospitality of the Center 
for Gravitational Wave Astronomy (UTB).  SLL also acknowledges support from
the Center for Gravitational Wave Physics, funded by the NSF under
cooperative agreement PHY 01-14375, and from NASA award NNG05GF71G. AJR
acknowledges the support of Sigma Xi and the hospitality of the
Nicolaus Copernicus Astronomical Center.  The majority of AJR's
calculations for this work were carried out at New Mexico State
University and the Harvard-Smithsonian Center for Astrophysics.  
The authors also would like
to thank Sam Finn for directing us to the KDE package for Matlab which was used to
generate the PDFs of the various channels, and Joe Romano for
providing a routine for generating the {\em LISA} noise.  Finally, we thank
the referee Gijs Nelemans and the anonymous referee for highly insightful questions 
and comments.  {\tt StarTrack} simulations were performed at the Copernicus 
Center in Warsaw, Poland.

\clearpage
 
\begin{deluxetable}{ccc}
\tablewidth{450pt}
\tablecaption{Detached Disc double WD Formation Channels\tablenotemark{a}}
\tablehead{Channel\tablenotemark{b} & Evolutionary History\tablenotemark{c} & rel. to Disc\tablenotemark{d}}
\startdata
HeHe-D1 & MT1(3-1) CE2(10-3;10-10)                         &     8\%     \\
HeCO-D1 & MT1(3-1/2/3) CE2(10-5;10-8)                      &     6\%     \\
COCO-D1 & MT1(2/3/4-1) CE2(4/7-5;7-8) MT2(7-8)             &     7\%     \\
COCO-D2 & MT1(2/3/4-1) CE2(11-5;11-8)                      &     3\%     \\
COCO-D3 & MT1(2/3/4-1/2/3) CE2(7-5;7-8)                    &     1\%     \\
COCO-D4 & MT1(2/3-1) MT1(8/9-1) CE2(11-3;11-7) MT2(11-8)   &     1\%     \\
COCO-D5 & MT1(3/4-1) CE2(7-3;7-7) MT2(7-8) MT1(8-11)       & $\lesssim$ 1\% \\
COHe-D1 & CE1(6-1;11-1) CE2(11-3;11-10)                    &     5\%     \\
the rest & ---                                            &      4\%     \\
\hline  
total &  $8.6 \times 10^{6}$  &    35\%     \\
\hline
\enddata
\label{channelsDD}
\tablenotetext{a}{GR freq. range $1\times10^{-4} - 0.01$ Hz ($5.6$ hrs $- 200$ s orbital periods).}  
\tablenotetext{b}{In channel names $XXYY$, $XX$ represents the first-formed WD.}
\tablenotetext{c}{CE -- common envelope phase; MT -- mass transfer (RLOF) phase.     
Numbers following CE or MT indicate the binary component donor star, either 1 
(initially more massive star on the zero-age MS [ZAMS]) or 2 (initially less 
massive star).  The numbers in parenthesis correspond to the donor and accretor 
types;  The following can be used as a guide \citep{Bel08}:\,
0: MS star $\leq 0.7 \msun$ \,
1: MS star $ > 0.7 \msun$ \, 
2: Hertzsprung gap \, 
3: red giant \,
4: core helium burning \,
5: early AGB \,
6: thermally pulsating AGB \,
7: MS naked helium star \,
8: Hertzsprung gap naked helium star \,
9: giant branch naked helium star \,
10: helium WD \,
11: carbon-oxygen WD \,
12: oxygen-neon WD \,
16: hydrogen WD \,
17: hybrid WD. \,
In the CE phase, the stars prior to and after the semi-colon 
represent the stages of evolution at the onset of and following the CE
event, respectively.}
\tablenotetext{d}{Percentage is relative to {\em LISA} binaries in said component (Disc or Bulge).}
\end{deluxetable}

\begin{deluxetable}{ccc}
\tablewidth{450pt}
\tablecaption{RLOF disc double WD Formation Channels\tablenotemark{a}}
\tablehead{Channel & Evolutionary History & rel. to Disc}
\startdata
COHe-R1 & CE1(5/6-1;8/11-1) CE2(11-3;11-10) MT2(11-10)      &   47\%     \\
COHyb-R1 & CE1(5/6-1;8/11-1) CE2(11-3;11-7/17) MT2(11-7/17) &   8\%     \\  
HeCO-R1 & MT1(2/3-1) CE2(10-5;10-8) MT1(10-11)              &   5\%       \\  
the rest   & ---                                            &   5\% \\
\hline
total      & $ 16.2 \times 10^{6}$     &    65\% \\ 
\hline
\enddata
\label{channelsDR}
\tablenotetext{a}{Tablenotes are the same as in Table 1.}
\end{deluxetable}

\clearpage 

\begin{deluxetable}{ccc}
\tablewidth{450pt}
\tablecaption{Detached bulge double WD Formation Channels\tablenotemark{a}}
\tablehead{Channel & Evolutionary History & rel. to Bulge}
\startdata
HeHe-D1 & MT1(3-1) CE2(10-3;10-10)                         &  20\% \\
COHe-D1 & CE1(6-1;11-1) CE2(11-3;11-10)                    &  9\% \\
HeCO-D1 & MT1(3-1/2/3) CE2(10-5;10-8)                      &  5\%  \\
COCO-D & various channels                                & 3\% \\
the rest   & ---                                           & 5\% \\
\hline
total      & $ 3.9 \times 10^{6}$             &  42\% \\ 
\hline
\enddata
\label{channelsBR}
\tablenotetext{a}{Tablenotes are the same as in Table 1.}
\end{deluxetable}

\begin{deluxetable}{ccc}
\tablewidth{450pt}
\tablecaption{RLOF Bulge double WD Formation Channels\tablenotemark{a}}
\tablehead{Channel & Evolutionary History &  rel. to Bulge}
\startdata
COHe-R1 & CE1(5/6-1;8/11-1) CE2(11-3;11-10) MT2(11-10)      & 37\%  \\
COHe-R2 & CE1(5-1;8-1) MT1(8-1) MT2(11-1/0) MT2(11-10)      & 20\%  \\
the rest & ---                                             &  1\%     \\
\hline  
total &   $ 5.4 \times 10^{6}$    &    58\%     \\
\hline
\enddata
\label{channelsBD}
\tablenotetext{a}{Tablenotes are the same as in Table 1.}
\end{deluxetable}

\clearpage 

\begin{deluxetable}{lcc}
\tablewidth{450pt}
\tablecaption{Expected number of resolvable {\em LISA} double WDs}
\tablehead{Type\tablenotemark{a} &  \% SNR $> 10$  & \% SNR $> 5$ }
\startdata
Detached & & \\
\hspace*{0.5cm} HeHe-D1  & 8.4    & 9.4   \\
\hspace*{0.5cm} HeCO-D1  & 6.3    & 5.5   \\
\hspace*{0.5cm} COCO-D1  & 1.2    & 1.8     \\ 
\hspace*{0.5cm} COCO-D2  & 3.6    & 3.2     \\
\hspace*{0.5cm} COCO-D3  & 2.3    & 1.8     \\
\hspace*{0.5cm} COCO-D4  & 9.1    & 5.3   \\
\hspace*{0.5cm} COCO-D5  & 1.4    & 1.4      \\
\hspace*{0.5cm} COHe-D1  & 23.0   & 22.3   \\
\hspace*{0.5cm} COHe-D2  &  0     & $< 1$   \\
\hspace*{0.5cm} COHyb-D1 &  1.2   & 1.2      \\
\hspace*{0.5cm} the rest &  $< 1$ &  $< 1$   \\
\hspace*{0.5cm} total    &  57.0  & 53.0  \\
RLOF & & \\
\hspace*{0.5cm} COHe-R1   &  17.3 & 21.4   \\
\hspace*{0.5cm} COCO-R1   &  10.2  & 11.0   \\ 
\hspace*{0.5cm} COHyb-R1  &  9.9 & 8.4   \\
\hspace*{0.5cm} HeCO-R1   &  2.6  &  2.6   \\
\hspace*{0.5cm} HeHe-R1   &  1.6  &  1.9   \\
\hspace*{0.5cm} HybCO-R1  &  1.2  &  1.5   \\
\hspace*{0.5cm} the rest  &  $ < 1$&   $< 1$  \\
\hspace*{0.5cm} total     &  43.0 & 47.0   \\
\hline & & \\
\hspace*{0.5cm} TOTAL     &  $2800$  & $11300$   \\
\enddata
\label{res}
\tablenotetext{a}{Formation channels producing resolvable {\em LISA} double WDs in the 
Galaxy, assuming that every binary with SNR $\geq 10$ or SNR $\geq 5$ with 
a gravitational wave frequency below 0.01 Hz has a chance to be 
resolved.  Relative percentages (with respect to the total resolved
population for the appropriate SNR threshold) for each channel are indicated.}
\end{deluxetable}
\clearpage

\end{document}